\documentclass[12pt]{article}
\usepackage{amsmath}
\usepackage{amssymb}

\usepackage{graphicx,psfrag,epsf}
\usepackage[round]{natbib}
\usepackage{url} % not crucial - just used below for the URL
\usepackage{comment}

\newcommand{\blind}{0}

\usepackage{subfig}
\usepackage{algorithm,algpseudocode,etoolbox}
\usepackage{booktabs}
\usepackage{psfrag,epsf}
\usepackage{caption}
\usepackage{xcolor}

\AtBeginEnvironment{algorithmic}{\let\textbf\relax}

\newtheorem{theorem}{Theorem}
\newtheorem{corollary}{Corollary}[theorem]

\begin{document}

\def\spacingset#1{\renewcommand{\baselinestretch}%
{#1}\small\normalsize} \spacingset{1}

\date{Sep 11, 2024}

%%%%%%%%%%%%%%%%%%%%%%%%%%%%%%%%%%%%%%%%%%%%%%%%%%%%%%%%%%%%%%%%%%%%%%%%%%%%%%

\if0\blind
{
  \title{\bf Sparsified Simultaneous Confidence Intervals for High-Dimensional Linear Models}
  \author{Xiaorui Zhu, Yichen Qin, and Peng Wang\thanks{Xiaorui Zhu is an Assistant Professor in the Department of Business Analytics \& Technology Management, Towson University. Yichen Qin is a Professor in the Department of Operations, Business Analytics, and Information Systems at the University of Cincinnati. Peng Wang is an Associate Professor in the Department of Operations, Business Analytics, and Information Systems at the University of Cincinnati.} 
}
  \maketitle
} \fi

\if1\blind
{
  \bigskip
  \bigskip
  \bigskip
  \begin{center}
    {\LARGE\bf Sparsified Simultaneous Confidence Intervals for High-Dimensional Linear Models}
\end{center}
  \medskip
} \fi

\bigskip
\begin{abstract}
Statistical inference of the high-dimensional regression coefficients is challenging because the uncertainty introduced by the model selection procedure is hard to account for. Currently, the inference of the model and the inference of the coefficients are separately sought. A critical question remains unsettled; that is, is it possible to embed the inference of the model into the simultaneous inference of the coefficients? If so, then how to properly design a simultaneous inference tool with desired properties? To this end, we propose a notion of simultaneous confidence intervals called the sparsified simultaneous confidence intervals (SSCI). Our intervals are sparse in the sense that some of the intervals' upper and lower bounds are shrunken to zero (i.e., $[0,0]$), indicating the unimportance of the corresponding covariates. These covariates should be excluded from the final model. The rest of the intervals, either containing zero (e.g., $[-1,1]$ or $[0,1]$) or not containing zero (e.g., $[2,3]$), indicate the plausible and significant covariates, respectively. The SSCI intuitively suggests a lower-bound model with significant covariates only and an upper-bound model with plausible and significant covariates. The proposed method can be coupled with various selection procedures, making it ideal for comparing their uncertainty. For the proposed method, we establish desirable asymptotic properties, develop intuitive graphical tools for visualization, and justify its superior performance through simulation and real data analysis.

\end{abstract}

\noindent%
\vfill

\newpage
\spacingset{1.5} % DON'T change the spacing!
\section{Introduction}
\label{sec:intro}

High-dimensional data analysis plays an important role in modern scientific discoveries. There has been extensive work on high-dimensional variable selection and estimation using penalized regressions, such as Lasso \citep{tibshirani_regression_1996}, SCAD \citep{fan_variable_2001}, MCP \citep{zhang_regularization_2010}, adaptive Lasso \citep{zou_adaptive_2006}, 
%nonconvex constrained $\ell_0$ likelihood \citep{shen_likelihood_based_2012}, 
and selection by partitioning solution paths \citep{liu_selection_2018}. 
In recent years, inference for the true regression coefficients and the true model began to attract attention. 
A major challenge of high-dimensional inference is how to quantify the uncertainty of the coefficient estimate because such uncertainty depends on two components, the uncertainty in selecting the model, and the uncertainty in parameter estimation given the selected model, 
both of which are difficult to estimate and are actively studied.

For inference of the regression coefficients, \citet{scheffe_method_1953} introduces the notion of simultaneous confidence intervals, which is a sequence of intervals containing the true coefficients at a given probability. 
For the high-dimensional linear models, \citet{dezeure_high-dimensional_2017} and \citet{zhang_simultaneous_2017} construct the simultaneous confidence intervals by bootstrapping the debiased Lasso approach \citep{geer_asymptotically_2014,zhang_confidence_2014}, and \citet{belloni_high-dimensional_2022} applies the similar idea in the context of linear instrumental variable model. 
\citet{yuan2022high} also utilizes the debiased Lasso to develop an inference tool.
Another route to achieve this objective is the simultaneous confidence region \citep{chatterjee_bootstrapping_2011,javanmard_confidence_2014}, but the boundary of the simultaneous confidence region is a function of all coefficients, making it hard to interpret and visualize. 
There is also a parallel stream of research in post-selection intervals for regression coefficients for high-dimensional linear models \citep{lee_exact_2016, lockhart_significance_2014, tibshirani_exact_2016}. 
However, their focus is on the coefficients in the selected model particularly, 
while our target is the entire regression coefficients with the consideration of the selection uncertainty. 
Besides, \citet{liu_bootstrap_2020} introduces bootstrap lasso + partial ridge estimator to construct the individual confidence interval, which does not focus on the simultaneous confidence intervals. 
Other inference tools include the multiple simultaneous testing proposed by \citet{ma_global_2021} and the adaptive confidence interval in a study of \citet{cai_confidence_2017}.
Many of the aforementioned simultaneous confidence intervals are often too wide and have non-zero widths for all covariates regardless of their significance. So the estimation uncertainty cannot be efficiently reflected.

For the inference of the true model, \citet{ferrari_confidence_2015} introduces variable selection confidence set by constructing a set of models that contain the true model with a given confidence level \citep{Zheng2019a, Zheng2019b}. 
Through the size of the model set, this inference tool characterizes the model selection uncertainty. 
\citet{li2019model} propose to construct the
lower bound model and upper bound model such that the true model is trapped in between at
a pre-specified confidence level \citep{wang_CG_2021, qin_visualization_2020}.
\citet{hansen_model_2011} propose to use a hierarchical
testing procedure to form the model confidence set for prediction.
Besides, \citet{nan_variable_2014} propose variable selection deviation measures for the quantification of model selection uncertainty.
Although these methods show promise in conducting inference for the true model in low-dimensional cases, their performances for high-dimensional models are often unsatisfactory.

Given the importance of inference for both the true regression coefficients and the true model, a critical question remains unsettled; that is, is it possible and how to embed the inference of the model into the inference of the parameters? To address this challenge, we propose a new notion of simultaneous confidence intervals, termed sparsified simultaneous confidence intervals.
Specifically, we construct a sequence of confidence intervals $\{ [ \underline{{\beta}}_j , \overline{{\beta}}_j] \}_{j=1}^p$ which simultaneously contain the true coefficients with $1-\alpha$ confidence level. 

Our simultaneous confidence intervals are sparse in the sense that some of its intervals' upper and lower bounds are shrunken to zero (i.e., $[0,0]$), meaning that the corresponding covariates are unimportant and should be excluded from the final model. The other intervals, either containing zero (e.g., $[-1,1]$ or $[0,1]$) or not containing zero (e.g., $[2,3]$), classify the corresponding covariates into the plausible ones and significant ones, respectively. The plausible covariates are weakly associated with the response variable. The significant covariates are strongly associated with the response variable and should be included in the final model. Therefore, the proposed intervals offer more information about the true coefficient vector than classical non-sparse simultaneous confidence intervals. In addition, the proposed method naturally provides two nested models, a lower bound model (that includes all the significant covariates) and an upper bound model (that includes all the significant and plausible covariates), so that the true model is trapped in between them at the same confidence level.

\begin{figure}[h]
\centering
\subfloat[Sparsified simultaneous confidence intervals.\label{fig:Grap_a}]{\includegraphics[width=0.9\textwidth]{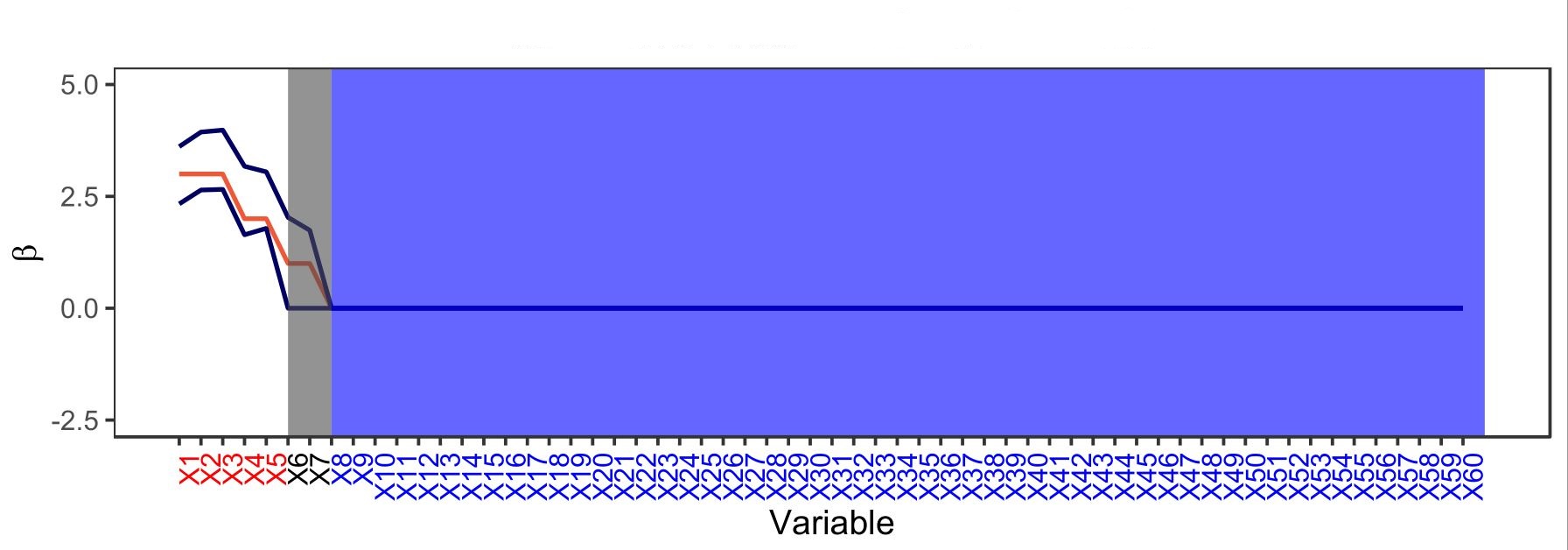}}\hfill
\subfloat[Simultaneous confidence intervals by debiased Lasso.\label{fig:Grap_b}] {\includegraphics[width=0.9\textwidth]{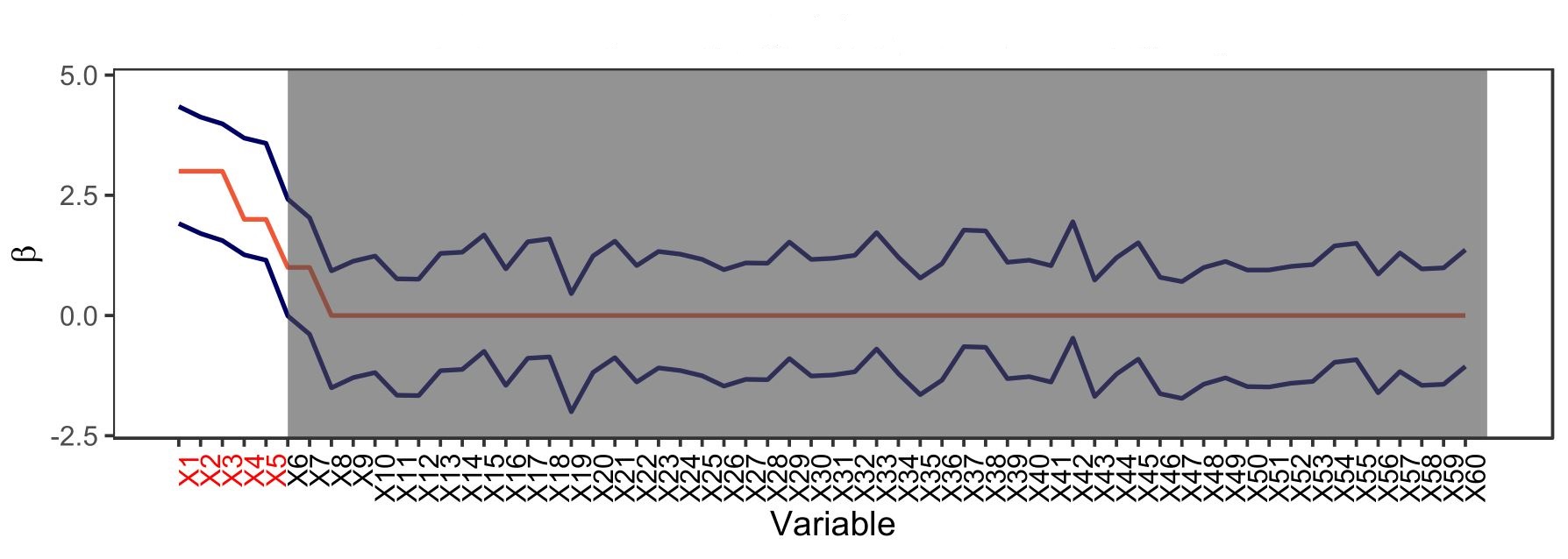}}\hfill
\caption{Graphical comparison of our method and \citet{zhang_simultaneous_2017}'s method.} \label{fig:Graphical}
\end{figure}

To illustrate the practical advantage of our SSCI over the existing simultaneous confidence intervals, we present an example to compare our method to \citet{zhang_simultaneous_2017}'s method, simultaneous confidence interval by debiased Lasso. We simulate 50 observations from the linear model $\mathbf{y} = \mathbf{X} \boldsymbol{\beta}^{0} + \boldsymbol{\varepsilon}$ where $\boldsymbol{\beta}^{0}={(3,3,3,2,2,1,1 ,0 ,...,0)}^T \in \mathbb{R}^{60}$ and both covariates and random error are standard normal. 
We construct both types of confidence intervals at 95\% confidence level in Figure \ref{fig:Graphical}. The true coefficient vector is in red, and the upper bound and lower bound of the simultaneous confidence intervals are colored in dark blue. Although both methods contain the true coefficient vector, our method is significantly narrower and presents more insights into the true coefficients and model. For example, in the blue-shaded area, the intervals' upper and lower bounds are shrunken to exact zero (i.e., $[0,0]$), which implies that these unimportant covariates should be excluded from the final model. The plausible covariates are plotted in a grey-shaded area, for which we do not have enough evidence to decide to include or exclude in the model. However, by the signs of the intervals, we at least can infer they all have positive or zero effects. The significant covariates in red labels should be included in the model. In contrast, \citet{zhang_simultaneous_2017} 's method carries limited model information since all the intervals in the grey-shaded area contain zero. It does not yield any inference about whether these covariates should be included or excluded in the true model. Meanwhile, for the simultaneous confidence interval based on the debiased Lasso, the widths of all intervals are the same. So it ignores the difference in estimation uncertainty.

To summarize, this article contributes to the literature in the following aspects. 
We integrate the sparsity feature into the simultaneous confidence intervals to provide insights on model selection uncertainty besides carrying out inference for regression coefficients.
Under a consistent model selection procedure, we have established the asymptotic coverage probability.
We develop a graphical representation of the proposed method to enhance the visualization for both parameter and model uncertainty.
In addition to satisfactory performance in typical high-dimensional settings, we numerically show that the proposed simultaneous confidence intervals perform well in weak signal settings compared to the existing methods.

% Contributions: 
% 1. Sparsity of simultaneous confidence intervals;
% 2. (Somewhere else) difference with post-selection inference; (post-selection inference in Lee et al. (2016) and the serial papers.)
% 3. Graphical tool to visualize the model selection uncertainty;
% 4. We develop asymptotic theories of SSCI under a consistent model selection procedure ();
% 5. Keep the weak signal contribution, but numerically show our SSCI performs reasonably well. Separate the simulation study 2 to a subsection. 

% Literature review:
% Rong Ma, T. Tony Cai & Hongzhe Li(2021) propose a global null hypothesis and simultaneous multiple testing. Only test statistics are developed and they said it is also of interest to construct confidence intervals but they did not do it. 
% T. T. Cai and Z. Guo (2017) propose a minimax length (or adaptive) of confidence intervals that also have the same length for each coefficient.
% Adel Javanmard and Andrea Montanari (2014) propose the Simultaneous Confidence Intervals but it is a confidence region, which means it has length for every coefficient in the set. Although the set could kick out some coefficients, the coverage probability is not valid for the coefficients excluding from the set. 

The article is organized as follows. In Section \ref{sec:SSCI}, we introduce our sparsified simultaneous confidence intervals, establish its theoretical properties, develop its graphical presentation, and discuss its connections to other methods. We explain the key ingredient in our approach, selection by partitioning solution paths, in Section \ref{sec:SPSP}. Numerical experiments and real data examples evidence the advantages of our approach in Sections \ref{sec:simulation} and \ref{sec:real-data-examples}. We conclude in Section \ref{sec:conc} and relegate proofs to the supplementary materials.

\section{Inference for High-Dimensional Linear Models}
\label{sec:SSCI}

Throughout the article, we focus on the inference tasks for the linear model $\mathbf{y} = \mathbf{X} \boldsymbol{\beta}^{0} + \boldsymbol{\varepsilon}$ where $\boldsymbol{\varepsilon}\sim N(\mathbf{0},  \sigma^2\mathbf{I}_n)$, $\mathbf{y} \in \mathbb{R}^n$ is the response vector, and $\mathbf{X} \in \mathbb{R}^{n \times p}$ is the fixed design matrix containing $p$ covariates. The parameter vector $\boldsymbol{\beta}^{0}=(\beta^{0}_1, \cdots, \beta^{0}_p)^T \in \mathbb{R}^p$ is assumed to be sparse with a small number of nonzero coefficients. We denote the index sets of the nonzero and zero coefficients as $\mathcal{S}_{0}=\{j: \beta^{0}_j\ne0\}$ and $\mathcal{S}_{0}^{c}=\{j: \beta^{0}_j=0\}$, respectively. We denote the cardinality of the active set as $s_0 = |\mathcal{S}_{0}|$ and assume $1 \le s_0 < p$, and the dimension $p$ can be larger than $n$.

%%%%%%%%%%%%%%%%%%%%%%%%%%%%%%%%%%%%%%%%%%%%%%%%%%%%%%%%
\subsection{Sparsified Simultaneous Confidence Intervals}\label{s:Sparsified-SCI} 
%%%%%%%%%%%%%%%%%%%%%%%%%%%%%%%%%%%%%%%%%%%%%%%%%%%%%%%%

We propose a new type of simultaneous confidence intervals, namely sparsified simultaneous confidence intervals (SSCI). It consists of a sequence of confidence intervals that contains all the true coefficients simultaneously with a confidence level $1-\alpha$. That is, let $\text{SSCI}_{1-\alpha} = \{\boldsymbol{\beta} \in \mathbb{R}^p : \underline{{\beta}}_j \le \beta_j \le \overline{{\beta}}_j, j=1,\dots,p \}$ where $\underline{{\beta}}_j$ and $\overline{{\beta}}_j$ are the lower and upper bounds of $j$-th coefficient, such that $\mathbf{P}(\boldsymbol{\beta}^0 \in \text{SSCI}_{1-\alpha}) = 1-\alpha$. 
The proposed SSCI is sparse in the sense that some of its intervals' upper and/or lower bounds are shrunken to zero, signaling the significance of the corresponding coefficients. 

To construct SSCI, we bootstrap a two-stage estimator that contains both the consistent model selection procedure and the refitted estimation procedure. Given an observed sample, we first apply the two-stage estimator to select an initial model and obtain the refitted coefficient estimate. We further generate bootstrap samples using the refitted estimate via the celebrated residual bootstrap \citep{freedman_bootstrapping_1981, efron_bootstrap_1979}.  
For each bootstrap sample, we apply the same two-stage estimator to identify the bootstrap model $\hat{\mathcal{S}}^{(b)}$ and obtain the refitted bootstrap coefficient estimate $\hat{\boldsymbol{\beta}}^{(b)}$. After removing the $\alpha$ proportion of bootstrap estimates according to their outlyingness, we use the remaining bootstrap estimates to construct SSCI. The details of this procedure are outlined in Algorithm \ref{alg:algorithm1}.

\begin{algorithm}[t]
	\caption{Sparsified simultaneous confidence intervals}\label{alg:algorithm1}
	\vspace{-3mm}
		\begin{itemize}\setlength{\itemindent}{.35in}
		    \setlength\itemsep{-0.5em}
			\item[Input: ]  $\mathbf{y}$, $\mathbf{X}$;
			\item[Output: ]  Sparsified simultaneous confidence intervals at the confidence level of $1-\alpha$;
			
			\item[Step 1:] Apply a two-stage estimation procedure on $\left\{\mathbf{y},\mathbf{X}\right\}$ to obtain the selected model $\tilde{\mathcal{S}}_{0}$ and its refitted coefficient estimate $\tilde{\boldsymbol{\beta}}$; 
			
			\item[Step 2:] Generate bootstrap samples $\left\{(\mathbf{y}^{(b)},\mathbf{X})\right\}^B_{b=1}$ via residual bootstrap with $\tilde{\boldsymbol{\beta}}$;
			
			\item[Step 3:] For $b=1, \dots, B,$ apply the same two-stage estimation procedure on $(\mathbf{y}^{(b)},\mathbf{X})$ to obtain bootstrap model $\hat{\mathcal{S}}^{(b)}$ and its refitted bootstrap estimate $\hat{\boldsymbol{\beta}}^{(b)}$;
			
			\item[Step 4:] Compute the outlyingness score $O^{(b)} =  \max_{j\in \{1,\dots,p\}} \big|(\hat{\beta}^{(b)}_j - \bar{\hat{\beta}}_j)/\widehat{\text{SE}}(\hat{\beta}_j)\big|, b=1,...,B$;
			
			\item[Step 5:] Construct the $\text{SSCI}_{1-\alpha}$ by letting   $\underline{{\beta}}_j=\min_{b\in \mathcal{B}_{1-\alpha}} \hat{\beta}^{(b)}_j$ and $\overline{{\beta}}_j=\max_{b\in \mathcal{B}_{1-\alpha}} \hat{\beta}^{(b)}_j$ \\ \qquad \qquad for $j=1,...,p$,  where $\mathcal{B}_{1-\alpha}=\{b: O^{(b)} \le q_{1-\alpha}\}$  and $q_{1-\alpha}$ is $(1-\alpha)$-quantile of $\{O^{(b)}\}_{b=1}^{B}$.
		\end{itemize}
		\vspace{-3mm}
\end{algorithm}

%$\begin{align*} \text{SSCI}_{1-\alpha} = \Big\{\boldsymbol{\beta} \in \mathbb{R}^p : \underset{b\in \mathcal{B}_{1-\alpha}}{\text{min}} \hat{\beta}^{(b)}_j \le \beta_j \le \underset{b\in \mathcal{B}_{1-\alpha}}{\text{max}} \hat{\beta}^{(b)}_j, j=1,\dots,p \Big\},\end{align*} $

For each bootstrap estimate, we measure its outlyingness relative to other bootstrap estimates, i.e., outlyingness score, by calculating the maximum absolute value of standardized coefficients
$O^{(b)}=O(\hat{\boldsymbol{\beta}}^{(b)}) = \max_{j\in \{1,\dots,p\}} \big|(\hat{\beta}^{(b)}_j - \bar{\hat\beta}_j)/\widehat{\text{SE}}(\hat\beta_j)\big|$,
where $\bar{\hat\beta}_j=\sum^B_{b=1}\hat{\beta}^{(b)}_j/B$ and $\widehat{\text{SE}}(\hat\beta_j)=\big(\sum^B_{b=1}\big(\hat{\beta}^{(b)}_j - \bar{\hat\beta}_j\big)^2/(B-1)\big)^{1/2}$ is the bootstrap standard error estimate.  Intuitively, standardized coefficients reveal the variability of each bootstrap coefficient estimate, while the maximum absolute value among all covariates measures the overall outlyingness of the bootstrap estimate and the bootstrap model. Hence, the outlyingness scores identify the extreme and rare bootstrap models and estimates. For example,  in one bootstrap iteration, if a two-stage estimator selects a particular covariate that has never been selected in other bootstrap iterations, the outlyingness score of this coefficient may be very large, which implies that this bootstrap estimate may be unusual and should be discarded. In Step 5 of Algorithm \ref{alg:algorithm1}, we remove these outlying bootstrap estimates by cutting the outlying scores at its $(1-\alpha)$-percentile and use the remaining bootstrap estimates to construct our intervals.

Specifically, the sparsity of the proposed SSCI depends on the stability and sparsity of the adopted selection method. Therefore, the general rule of thumb is to use a stable selection method with a low false positive rate. This is because a stable selection method would produce selection results that are reflective of the difference among bootstrap samples,  such that different bootstrap samples will not produce overly dramatically different sets of selected variables. And a low false positive rate ensures that the selection result from each bootstrap sample is sparse. Then the union of the selected variables from the retained bootstrap samples, after removing the 5\% most outlying ones, will also be sparse.  
To better classify the sparse intervals, 
we define three groups of covariates: {\it significant} covariates whose intervals do not contain zero, denoted as $\hat{\mathcal{S}}_{\text{1}}=\{j: \underline{{\beta}}_j\cdot \overline{{\beta}}_j > 0 \}$; {\it plausible} covariates whose intervals contain zero but has non-zero width, denoted as $\hat{\mathcal{S}}_{\text{2}}=\{j: \underline{{\beta}}_j \cdot \overline{{\beta}}_j \le 0 \text{ and } \underline{{\beta}}_j \neq \overline{{\beta}}_j \}$, and {\it unimportant} covariates whose intervals contain zero and have zero-width, denoted as  $\hat{\mathcal{S}}_{\text{3}}=\{j: \underline{{\beta}}_j = \overline{{\beta}}_j = 0 \}$. The significant covariates are the ones strongly associated with the response variable and should be included in the final model at $1-\alpha$ confidence level. The plausible covariates are the ones weakly associated with the response. We do not have enough evidence to prove their significance nor to disqualify them for the final model. The unimportant covariates are the superfluous predictors that should be excluded from the final model at $1-\alpha$ confidence level. Therefore, not only do we have the estimation uncertainties for each coefficient estimate, but we also obtain model information based on the sparsity feature of the intervals.

With the proposed approach, the algorithm can be coupled with the two-stage estimators that consist of a stable and consistent model selection procedure followed by a consistent refitted regression estimator. For the model selection stage, we consider SCAD \citep{fan_variable_2001}, MCP \citep{zhang_regularization_2010}, Adaptive Lasso \citep{zou_adaptive_2006}, and \citet{liu_selection_2018}'s SPSP method based on the solution path of SCAD (SPSP+SCAD), MCP (SPSP+MCP), Adaptive Lasso (SPSP+AdaLasso), or Lasso (SPSP+Lasso). The details of the SPSP method will be explained later in Section 3. 

It is worth noting that the adopted selection method for constructing the SSCI should be stable and have a low false positive rate when the sample size is not very large. Specifically, a stable selection method that the number of distinct models produced by the bootstrap samples is not overly large. Moreover, the selection method with a low false positive rate can avoid over-selecting too many irrelevant covariates. Otherwise, the SSCI will be less sparsified and have many ``plausible'' intervals that contain the zero (e.g., $[-1,1]$ or $[0,1]$). For example, the Lasso selection is very unstable and has a high false positive rate when the tuning parameter is sub-optimal for a finite sample \citep{zhao_model_2006, liu_selection_2018, feng2013consistent}. Therefore, we do not recommend incorporating the SSCI with the Lasso+CV since the coverage rate and the sparsity of the SSCI might not be good enough.  

Through numerical studies, among all the selection methods we adopted, we found that SPSP is the best model selector for constructing the proposed intervals in terms of both the sparsity and the coverage probability.
When compared with other classical simultaneous confidence intervals, our SSCI using the SPSP selector is noticeably narrower and it offers quite sparse intervals that imply more accurate information of relevant and irrelevant covariates and the true model. Therefore, we recommend \citet{liu_selection_2018}'s method as the default selection method for our proposed approach. For the refitted estimator $\tilde{\boldsymbol{\beta}}$ in the second stage, we adopt the least square refitted estimator throughout this article, that is,  $\tilde{\boldsymbol{\beta}}_{\tilde{S}} = (\mathbf{X}_{\tilde{S}}^T \mathbf{X}_{\tilde{S}})^{-1} \mathbf{X}_{\tilde{S}}^T \mathbf{Y}  \in \mathbb{R}^{|\tilde{S}|}$ and $\tilde{\boldsymbol{\beta}}_{\tilde{S}^c}=\mathbf{0} \in \mathbb{R}^{p-|\tilde{S}|}$.

Lastly, the proposed method allows alternative outlyingness scores to be implemented. By defining outlyingness scores differently, we are able to construct different inference tools such as individual confidence intervals, simultaneous confidence intervals, or model confidence set \citep{ferrari_confidence_2015}. We leave this topic for future work.  

%%%%%%%%%%%%%%%%%%%%%%%%%%%%%%%%%%%%%%%%%%%%%%%%%%%%%%%%
\subsection{Theoretical Properties}\label{s:properties} 
%%%%%%%%%%%%%%%%%%%%%%%%%%%%%%%%%%%%%%%%%%%%%%%%%%%%%%%%

In this section, we establish the theoretical properties of the proposed method, such as the coverage probability of the SSCI under various model selection methods. Without loss of generality, we assume $\mathbf{X}$ are standardized with zero mean and unit variance. Let $\mathbf{X}_{\mathcal{S}_{0}} \in \mathbb{R}^{n \times s_{0}}$ be the true signal covariates matrix and $s_{0}=|\mathcal{S}_{0}|$.

\begin{theorem} \label{theorem_SCI} Suppose the selection method selects the true model with the probability $1-2e^{-h(n)}$, that is $\mathbf{P}(\tilde{\mathcal{S}}=\mathcal{S}_{0})=1-2e^{-h(n)}$. Then, with $B=o(e^{h(n)})$, for any confidence level of $\alpha\in (0,1)$, the $\textup{SSCI}_{1-\alpha}$ constructed in Algorithm \ref{alg:algorithm1} has the asymptotic coverage probability  $\mathbf{P}(\boldsymbol{\beta}^0 \in \textup{SSCI}_{1-\alpha}) \xrightarrow{n\rightarrow\infty} 1-\alpha.$
\end{theorem}

\textbf{Remark 1.} Theorem \ref{theorem_SCI} implies that the proposed SSCI can asymptotically achieve the nominal coverage probability with a consistent selection procedure. 
The function $h(n)$ has different forms when different selection method is adopted. We have $h(n)=n^c$ with $0 \le c <1$ for Lasso, and $h(n)=\gamma n$ for Adaptive Lasso and SPSP. We further discuss the induced properties from Theorem \ref{theorem_SCI} when we incorporate Lasso, adaptive Lasso, and SPSP in our approach as shown in Corollaries \ref{corollary_Lasso}, \ref{corollary_AdaLasso} and \ref{corollary_SPSP}.

\begin{corollary}[Lasso]\label{corollary_Lasso}
Let $\tilde{\mathcal{S}}^{\textup{Lasso}}(\lambda_n)$ be the model selected by the Lasso with the tuning parameter $\lambda_n$ and $\textup{SSCI}^{\textup{Lasso}}_{1-\alpha} (\lambda_n)$ be constructed by Algorithm \ref{alg:algorithm1} using the Lasso with the tuning parameter $\lambda_n$ and the least square refitted estimate. Under the strong irrepresentable condition \citep{zhao_model_2006}, $\mathbf{P}(\tilde{\mathcal{S}}^{\textup{Lasso}}(\lambda_n)=\mathcal{S}_{0}) \ge 1-2e^{-n^c}$ if $\lambda_n/n \rightarrow 0$ and $\lambda_n/n^{(1+c)/2} \rightarrow \infty$ with $0\le c < 1$, and $B=o(e^{n^c})$ , we have
$\mathbf{P}(\boldsymbol{\beta}^0 \in \textup{SSCI}^{\textup{Lasso}}_{1-\alpha}(\lambda_n)) \xrightarrow{n\rightarrow\infty} 1-\alpha.$
\end{corollary}

\begin{corollary}[Adaptive Lasso]\label{corollary_AdaLasso}
Let $\tilde{\mathcal{S}}^{\textup{Ada}}(\lambda_n)$ be the model selected by the adaptive Lasso with the tuning parameter $\lambda_n$ and $\textup{SSCI}^{\textup{Ada}}_{1-\alpha} (\lambda_n)$ be constructed by Algorithm \ref{alg:algorithm1} using the adaptive Lasso with the tuning parameter $\lambda_n$ and the least square refitted estimate. Under the restricted eigenvalue condition \citep{geer_adaptive_2011},  $\mathbf{P}(\tilde{\mathcal{S}}^{\textup{Ada}}(\lambda_n)=\mathcal{S}_{0})\ge 1-2e^{-\gamma n}$ if  $\lambda_n=4\sigma\sqrt{2\gamma+ 2\log p/n }$ and  $\gamma\rightarrow 0$. Further if $B=o(e^{\gamma n})$, we have
$\mathbf{P}(\boldsymbol{\beta}^0 \in \textup{SSCI}^{\textup{Ada}}_{1-\alpha}(\lambda_n)) \xrightarrow{n\rightarrow\infty} 1-\alpha.$
\end{corollary}

\begin{corollary}[SPSP]\label{corollary_SPSP}
Let $\tilde{\mathcal{S}}^{\textup{SPSP}}$ be the model selected by SPSP and $\textup{SSCI}^{\textup{SPSP}}_{1-\alpha}$ be constructed by Algorithm \ref{alg:algorithm1} using the SPSP and the least square refitted estimate. Under the compatibility condition in \citet{buhlmann_statistics_2011}, and the weak identifiability condition in \citet{liu_selection_2018}, the SPSP can select the true model $\mathcal{S}_{0}$ over $\lambda\in [4\sigma\sqrt{2\gamma + {2\log p}/{n}},+\infty]$ when $\gamma \rightarrow 0$ with probability at least $1-2e^{-\gamma n}$, that is $\mathbf{P}(\tilde{\mathcal{S}}^{\textup{SPSP}}=\mathcal{S}_{0})\ge (1-2e^{-\gamma n})$. Further if $B=o(e^{\gamma n})$, we have
$\mathbf{P} (\boldsymbol{\beta}^0 \in \textup{SSCI}^{\textup{SPSP}}_{1-\alpha} ) \xrightarrow{n\rightarrow\infty} 1-\alpha.$
\end{corollary}

Remark 2. The proposed SSCI procedure relies on the selection consistency of the adopted model selection approach and, therefore requires conditions to ensure the selection consistency. This includes a so-called $\beta$-min condition, which requires the minimal signal to be of order $s_0\sqrt{\log p/n}$ \citep{liu_selection_2018, buhlmann_statistics_2011}. This enables the SSCI procedure to produce the most informative inference results by producing sparse simultaneous confidence intervals.  
On the contrary, the de-biased type of approach proposed by \citet{zhang_simultaneous_2017} does not require the minimal strength assumption for the true signals at the expense of sacrificing the lengths of the confidence intervals, even for those whose true values are 0. However, they do require stronger conditions on the design matrix $\bf X$ to consistently estimate the inverse of its covariance matrix. Moreover, the number of non-zero coefficients needs to be  $o(\sqrt{n/(\log(p))^3}),$ which is stronger than  $O(\sqrt{n/\log(p)})$ required by our SSCI procedure. The approach by \citet{dezeure_high-dimensional_2017} follows a similar idea but uses a different bootstrap. They still need to impose sparsity and eigenvalue conditions on $\bf X$, but allow $s_0$ to be $O(\sqrt{n/\log(p)}).$ 

The proposed SSCI relies on the selection consistency of the adopted model selection procedure. Fortunately, most of these model selection procedures retain their properties when normal errors are replaced with sub-Gaussian errors, i.e. when $E e^{\mathbf{x}'\varepsilon} \le e^{\sigma^2_1 ||\mathbf{x}||^2/2}, \forall \mathbf{x}$, provided $\sigma^2_1 < \sigma^2$ \citep{zhang_regularization_2010, genzel2022generic}. The validity of the SPSP procedure depends on the estimation error bounds from the penalty functions it employs, which are also retained. Therefore, the proposed SSCI procedure will also retain its property when sub-Gaussian errors are present. Moreover, in the context of generalized linear models, as long as the selection procedure adopted is consistent, the proposed SSCI procedure will still maintain its coverage asymptotically \citep{park20071, calcagno2010glmulti, hao_model_2018}.

%%%%%%%%%%%%%%%%%%%%%%%%%%%%%%%%%%%%%%%%%%%%%%%%%%%%%%%%
\subsection{Visualization and Comparison with Other Methods}\label{s:graphical-SCI} 
%%%%%%%%%%%%%%%%%%%%%%%%%%%%%%%%%%%%%%%%%%%%%%%%%%%%%%%%
In this section, we develop an intuitive graphical tool, namely the SSCI plot, to visualize the estimation uncertainty. Specifically, we plot the confidence intervals of all covariates side by side but rearrange them in the following order: We place the significant covariates with all positive (or all negative) interval boundaries on the left (or right) end of the horizontal axis. The plausible and unimportant covariates are placed in the middle with grey and blue shades. The bootstrap estimates contained in SSCI are drawn in a light blue line while the true coefficient (if known) is in red.

\begin{figure}[h]
	\centering \includegraphics[width=0.9\textwidth, height=0.6\textwidth]{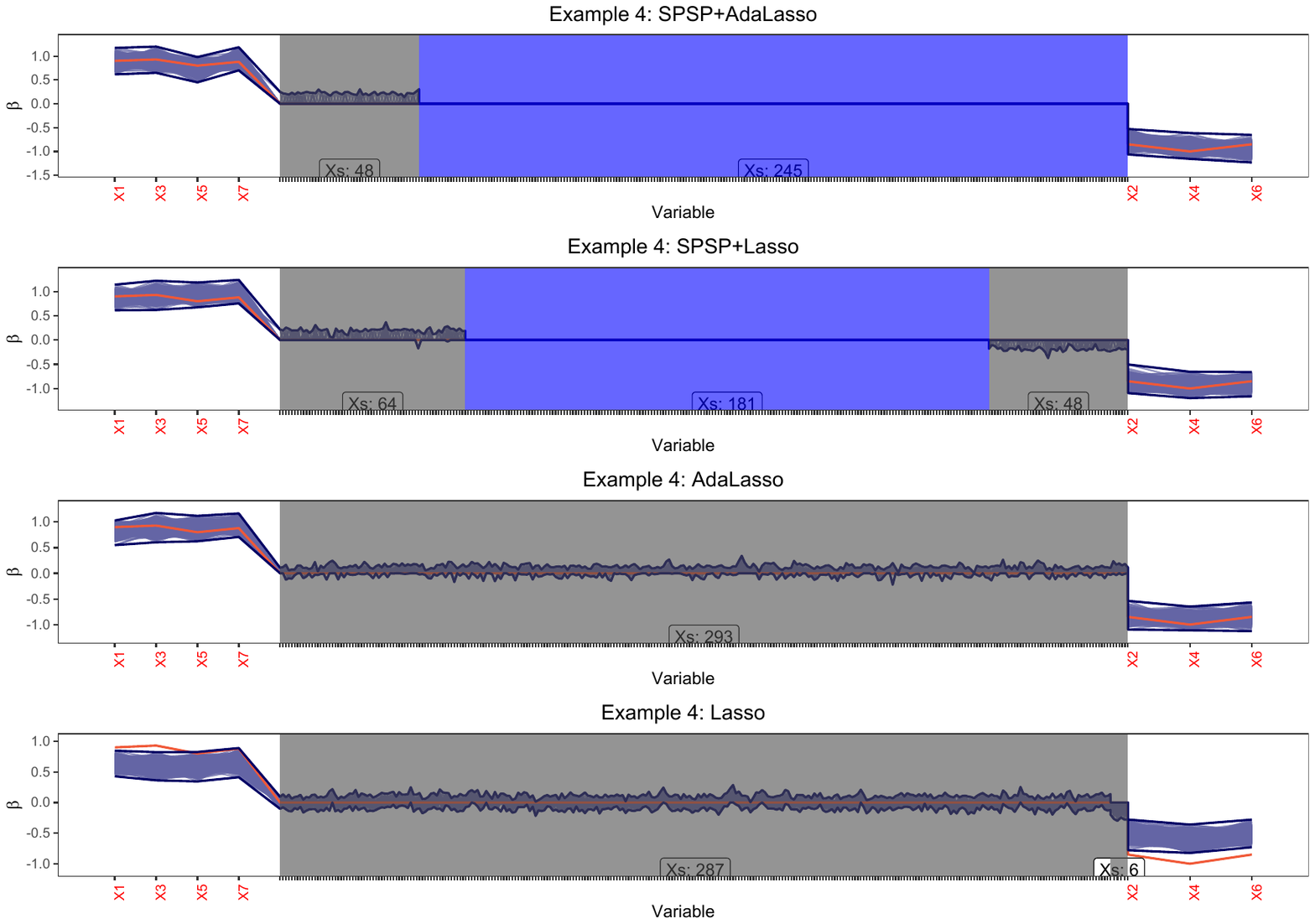} 
	\caption{Graphical presentation of SSCI.}\label{fig:Graphic_SSCI}
\end{figure}

We present a simple example using the simulated data from Example 4 in Section \ref{sec:simulation}. We construct SSCI using SPSP+AdaLasso, SPSP+Lasso, AdaLasso, and Lasso and visualize them in Figure \ref{fig:Graphic_SSCI}.
For the sake of simplicity, we only display the total number of covariates instead of the variables' names.

In the figure, the estimation uncertainty can be reflected by the vertical widths of the intervals. The significant covariates are labeled in red to highlight their importance in the model. For the plausible covariates, their confidence intervals contain zero; hence we put these covariates on the ``waiting list''. The confidence intervals of unimportant covariates all shrunk to zero, implying they should be excluded from the model. 

It is worth noting that the SSCIs by SPSP perform better than the others in this example. For instance, the plausible covariates of SSCI by SPSP are much fewer than the rest, indicating the stability of SPSP. This is because the bootstrap models of AdaLasso and Lasso fail to reach any consensus. In addition, the vertical widths of SSCIs by SPSP are, on average narrower than the other SSCIs, indicating a lower estimation uncertainty. We summarize their vertical widths in Table \ref{table:widthSSCI}.

\begin{table}[!h]
\normalsize
\centering
\caption{Comparison of the widths of SSCI by different selection methods}\label{table:widthSSCI}
\scalebox{1}{
\begin{tabular}{lcccc}
	\toprule
    & \multicolumn{4}{c}{Selection Method}  \\
	\cmidrule{2-5}  
 Width of SSCI &  SPSP+AdaLasso &  SPSP+Lasso &  AdaLasso &  Lasso \\ 
	\midrule
 Average width of all covariates &  0.048 &  0.086 &  0.183 &  0.226  \\
 Average width of signals        &  0.539 &  0.537 &  0.512 &  0.464  \\
 Average width of non-signals    &  0.036 &  0.076 &  0.175 &  0.220  \\
\bottomrule
\end{tabular}
}
\end{table}

Lastly, we graphically compare SSCI with the simultaneous confidence region (SCR) and the simultaneous confidence intervals based on debiased Lasso (SCI debiased Lasso), all of which capture the true coefficients at the $1-\alpha$ confidence level.

\begin{figure}
\centering
\subfloat[SCR\label{fig:SCR}]{\includegraphics[width=0.24\textwidth]{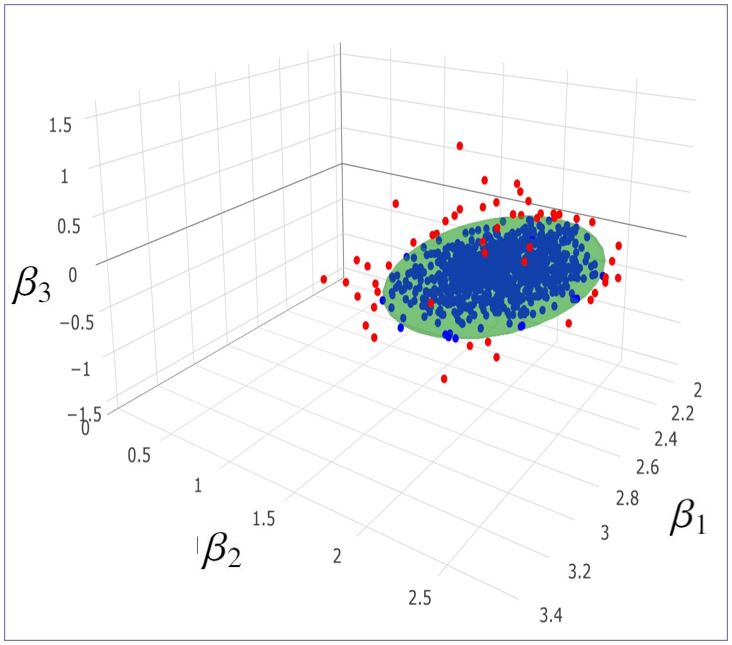}}
\subfloat[SCI \label{fig:SCI}] {\includegraphics[width=0.24\textwidth]{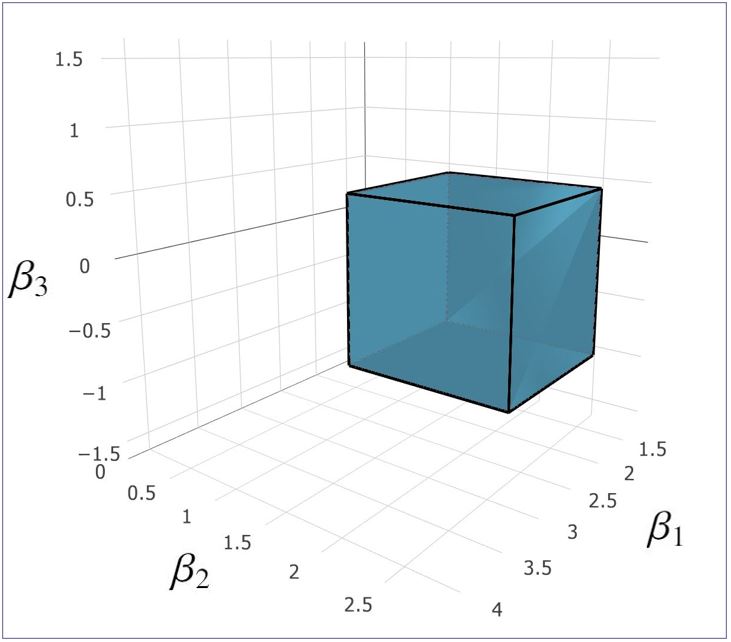}}
\subfloat[SSCI \label{fig:SSCI1}] {\includegraphics[width=0.24\textwidth]{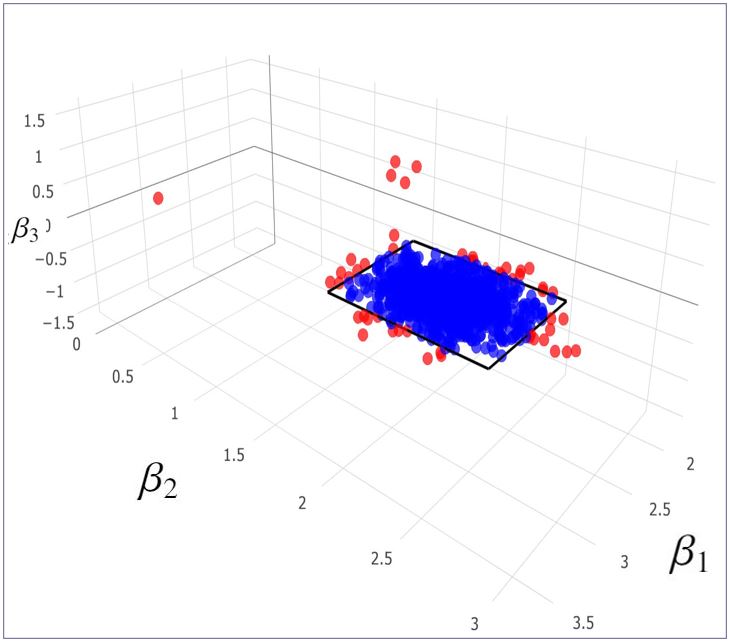}}
\subfloat[SSCI, 2nd view \label{fig:SSCI2}]{\includegraphics[width=0.24\textwidth]{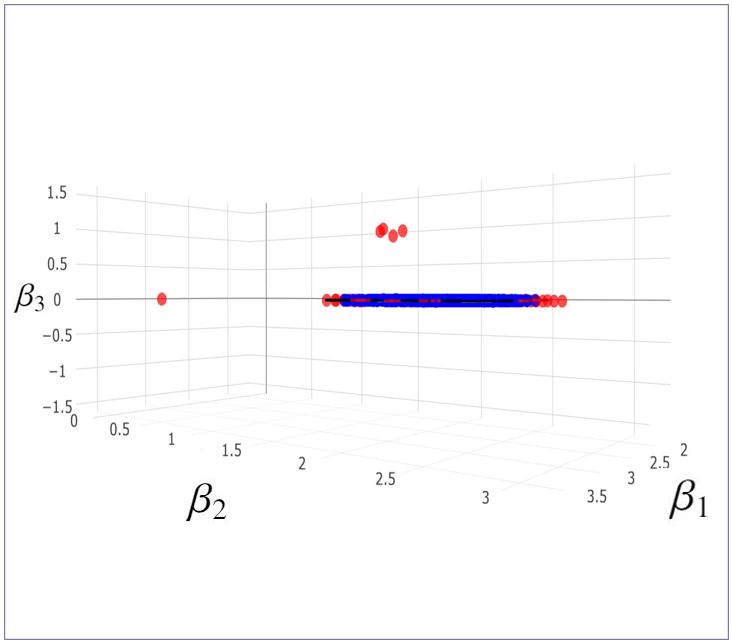}}
\caption{Graphical comparison of SCR, SCI debiased Lasso, and SSCI. Note that Panels (c) and (d) show the same SSCI from different perspectives.} 
\label{fig:GeoDifference2}
  
\end{figure}

The SCR \citep{chatterjee_bootstrapping_2011} is defined as  $\text{SCR}_{1-\alpha}=\{\boldsymbol{\beta}\in \mathbb{R}^p: \lVert \boldsymbol{\beta}-\hat{\boldsymbol{\beta}} \lVert\le n^{-1/2}\hat{t}_{1-\alpha}\}$,
where $\hat{\boldsymbol{\beta}}$ is the Lasso estimate, $\lVert \cdot \lVert$ is $\ell_2$ norm, and $\hat{t}_{1-\alpha}$ is the $1-\alpha$ quantile of the $\ell_2$ norms of the centered and scaled bootstrap estimates. Therefore, the shape of SCR is an ellipsoid centered at $\hat{\boldsymbol{\beta}}$ capturing $1-\alpha$ proportion of the bootstrap estimates, as shown in Figure \ref{fig:SCR}.
The data is simulated with $n=200$, $p=3$, $(\beta_1, \beta_2, \beta_3)=(3, 2, 0)$, $\sigma_{\varepsilon}=3.5$.

The SCI debiased Lasso \citep{zhang_simultaneous_2017} is defined as $
\text{SCI}^{\textup{deb}}_{1-\alpha}= \{\boldsymbol{\beta} \in \mathbb{R}^p: \sqrt{n} | \boldsymbol{\beta} - \hat{\boldsymbol{\beta}}^{\textup{deb}} | \le c^*_{1-\alpha}\mathbf{1} \}$ 
where $\hat{\boldsymbol{\beta}}^{\textup{deb}}$ is the debiased Lasso estimate, and $c^*_{1-\alpha}$ is $1-\alpha$ bootstrap critical value in \citet{zhang_simultaneous_2017}, and $\mathbf{1}=(1,...,1)^T$. 
Therefore, the shape of SCI debiased Lasso is a hypercube centered at $\hat{\boldsymbol{\beta}}^{\textup{deb}}$, as shown in Figure \ref{fig:SCI}, using the same data.

In contrast, our SSCI is a special type of hyperrectangle in $\mathbb{R}^p$ with some of its edges shrunken to zero, representing the unimportant covariates. An example is shown in Figures \ref{fig:SSCI1} and \ref{fig:SSCI2} using the same data. The bootstrapped SPSP estimates are shown in dots. We use red dots to represent the excluded bootstrap estimates when constructing the simultaneous confidence intervals. The blue dots are those bootstrap coefficient estimates after getting rid of the 5\% estimates on the tails. The SSCI is the black hyperrectangle with its third dimension, $\beta_3$, shrunken to zero.

Even though all three types of confidence sets claim to capture the true coefficients at the $1-\alpha$ confidence level, they are quite different in terms of usage. SCR often offers the tightest confidence set than SCI debiased Lasso since the ellipsoid is able to capture more bootstrap estimates than the hypercube of the same volume. SCI debiased Lasso is easy to interpret since it consists of $p$ individual intervals, whereas SCR cannot be expressed in the same fashion since the interval boundary for one coefficient depends on the rest of the coefficients. Unfortunately, neither case can offer us model information. On the other hand, SSCI is often one of the tightest due to the stability of SPSP, and it comes with easy interpretability. The shrunken edges of SSCI indicate the exclusion of the corresponding covariates in the final model. Such a message is not available in either SCR or SCI debiased Lasso.

%%%%%%%%%%%%%%%%%%%%%%%%%%%%%%%%%%%%%%%%%%%%%%%%%%%%%%%%
%\subsection{Decomposition of Uncertainty}\label{s:uncertainty} 
%%%%%%%%%%%%%%%%%%%%%%%%%%%%%%%%%%%%%%%%%%%%%%%%%%%%%%%%
%%%%%%%%%%%%%%%%%%%%%%%%%%%%%%%%%%%%%%%%%%%%%%%%%%%%%%%%
\subsection{Model Insights}\label{s:MCB} 
%%%%%%%%%%%%%%%%%%%%%%%%%%%%%%%%%%%%%%%%%%%%%%%%%%%%%%%%

Since the refitted parameter estimate $\tilde{\beta}_j$ is obtained after the model $\tilde{\mathcal{S}}$ is selected, the overall uncertainty of $\tilde{\beta}_j$, measured by SSCI width, is affected by two resources of the uncertainty, i.e., model selection uncertainty and estimation uncertainty conditional on the selected model.
However, the contributions of the estimation and model uncertainty to the width of SSCI cannot be easily or perfectly separated since these two sources of uncertainty are mingled
together and sequentially affect the interval width, i.e., first model selection uncertainty
and then estimation uncertainty conditional on the selected model.

With the consistent model selector and least square estimator, the estimation uncertainty, i.e., standard error, converges to 0 at the order $1/n$ \citep{casella2021statistical}.
Since the probability of selecting the true model is typical $1-2 \exp(-Cn)$, the model uncertainty converges to 0 much faster.
Hence, asymptotically, the overall estimation uncertainty is mostly the estimation uncertainty.
On the other hand, under the finite sample size, both the model uncertainty and estimation uncertainty are non-negligible.
In this case, we can still utilize SSCI to provide insights about models, admitting that the asymptotic results are not directly applicable and the finite sample theoretical results are hard to derive.

Under the finite sample size, SSCI naturally defines two nested models: 1) the lower bound model with the significant covariates, denoted as $\underline{\mathcal{S}}=\hat{\mathcal{S}}_{1}$;  2) the upper bound model with the significant and plausible covariates, denoted as $\overline{\mathcal{S}} = \hat{\mathcal{S}}_{1} \cup \hat{\mathcal{S}}_{2}$. It is straightforward to show that the two models trap the true model $\mathcal{S}_{0}$ with at least the same confidence level. 
We call this pair of models as the model confidence bounds (MCB) induced by SSCI. 
Conceptually, our MCB extends the idea of the classical confidence interval for a population parameter to the case of model selection. The lower bound model is regarded as the most parsimonious model that cannot afford to lose one more covariate. In contrast, the upper bound model is viewed as the most complex model that cannot tolerate one extra covariate. 

We can define the width of MCB as $w=|\overline{\mathcal{S}} \setminus \underline{\mathcal{S}}| = |\hat{\mathcal{S}}_{2}|$, i.e., the number of plausible covariates. 
Similar to the width of the classical confidence interval, the MCB width can be potentially used as a measure of model selection stability. 
Since MCB can be coupled with various selection methods, we can compare their MCB widths at the same confidence level. 

\begin{figure}[t]
	\centering \includegraphics[width=0.9\textwidth,height=0.45\textwidth]{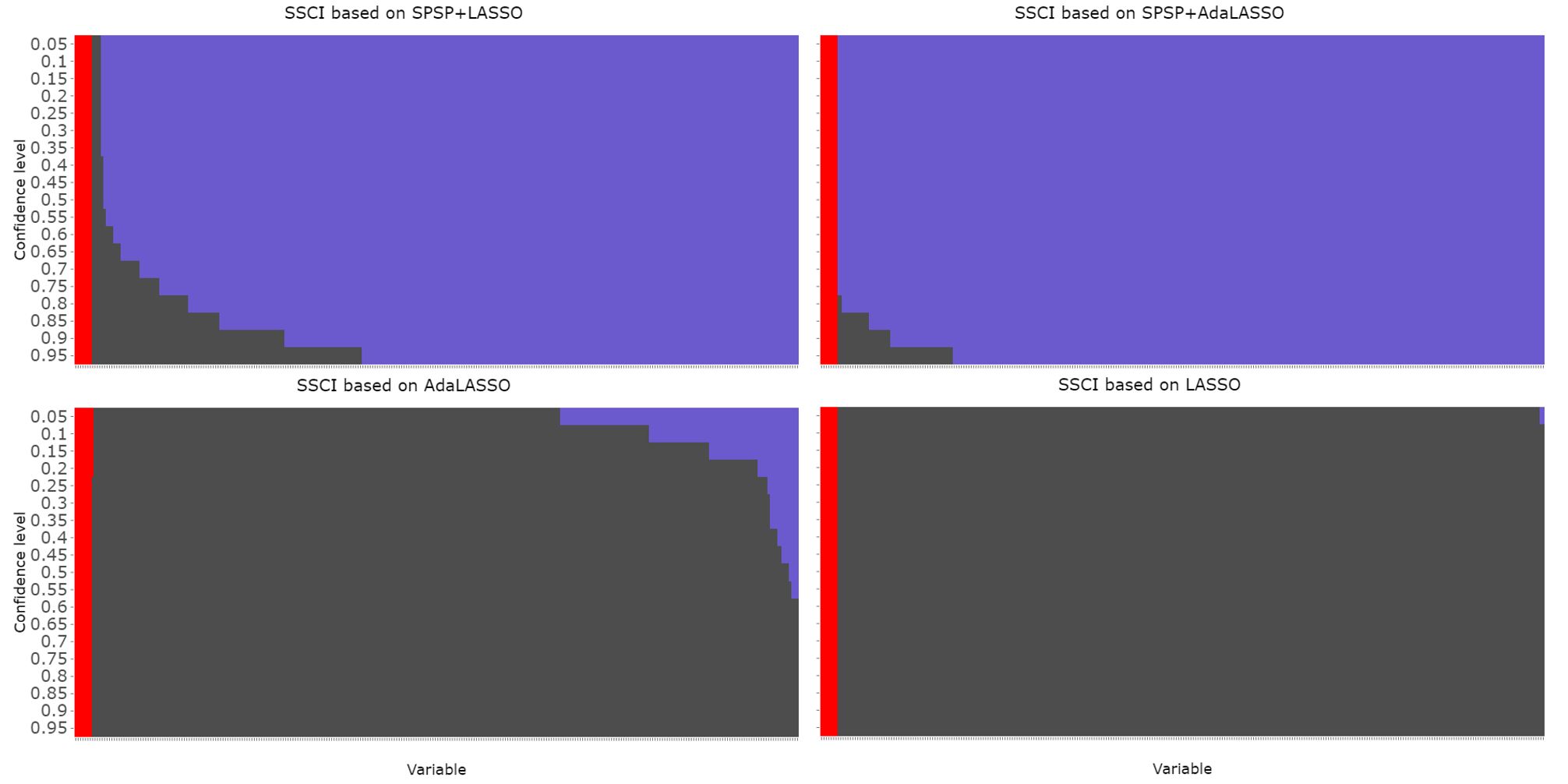}
	\caption{MCB induced by SSCI at various confidence levels by different selection methods.}\label{fig:Progression_SSCI}
\end{figure}

As an example, we construct the MCBs using SPSP+AdaLasso, SPSP+Lasso, AdaLasso, and Lasso at different confidence levels using the simulated data from Example 4 in Section \ref{sec:simulation}.  
Figure \ref{fig:Progression_SSCI} shows how these MCBs behave as the confidence level changes. The horizontal axis represents the covariate, whereas the vertical axis represents the confidence level. 
At each confidence level, the lower bound model (or the significant covariates) is the red area. 
The plausible covariates are the grey areas, which are also the MCB width for different confidence levels. The upper bound model consists of both red and grey areas. The unimportant covariates are the blue area. 

As we can see, the MCB width increases as the confidence level increases, which is consistent with the traditional confidence intervals.
Among these MCBs, the MCBs by SPSP are able to maintain small widths throughout the confidence levels, whereas the MCBs by AdaLasso and Lasso all have large widths. This evidence implies that the SPSP-based method selects the model more stably and results in fewer unique bootstrap models than Lasso and AdaLasso. 

\section{Selection by Partitioning Solution Paths}
\label{sec:SPSP}
Our SSCI can be equipped with different variable selection and estimation methods. A more accurate and stable method will help to construct narrower intervals as well as more zero-width intervals and provide an informative MCB. Throughout this article, we rely on SPSP \citep{liu_selection_2018} due to its stability and accuracy.

The idea of SPSP is to partition the covariates into ``relevant'' and ``irrelevant'' sets by utilizing the entire solution paths, rather than shrinking some coefficient estimates to zero given only one particular turning parameter. It mainly has three advantages. First, SPSP is more stable and accurate (low false positive and false negative rates) than other variable selection approaches because it uses information from the entire solution paths. As a result, it will generate fewer unique models among bootstrap samples. Second, SPSP is computationally more efficient because it does not need to calculate the solutions multiple times for all tuning parameters when cross-validation is involved. When the bootstrap technique is applied, this advantage could dramatically decrease the computing time. Third, one can flexibly incorporate SPSP with the solution paths of Lasso, adaptive Lasso, and other penalized estimators. 

Here, we briefly introduce the SPSP approach. A more in-depth discussion can be found in \citet{liu_selection_2018}. Given a tuning parameter $\lambda_k$ and the corresponding coefficients estimation $\hat{\beta}^{(k)}_{1}, \dots, \hat{\beta}^{(k)}_{p}$, it sorts the absolute values of the estimators in ascending order to create the order statistics $|\hat{\beta}^{(k)}|_{(1)}, \dots, |\hat{\beta}^{(k)}|_{(p)}$. Based on the order statistics, it defines the adjacent distances between the ordered values as $D^{(k)}_{j}= |\hat{\beta}^{(k)}|_{(j)} - |\hat{\beta}^{(k)}|_{(j-1)},  j = 1, \dots, p.$ The SPSP defines the gap between a relevant set ($\hat{\mathcal{S}}_k$) and an irrelevant set ($\hat{\mathcal{S}}^c_k$) as the adjacent distance between two order statistics, $|\hat{\beta}^{(k)}|_{(p-\hat{s}_k)}$ and $|\hat{\beta}^{(k)}|_{(p-\hat{s}_k+1)}$. It is $D(\hat{\mathcal{S}}_k, \hat{\mathcal{S}}^c_k)=D^{(k)}_{p-\hat{s}_k+1}= |\hat{\beta}^{(k)}|_{(p-\hat{s}_k+1)} - |\hat{\beta}^{(k)}|_{(p-\hat{s}_k)}$. With this notation, the largest adjacent distance in $\hat{\mathcal{S}}_k$ can be defined as $D_{\text{max}}(\hat{\mathcal{S}}_k)=\text{max}\{D^{(k)}_j: j>p-\hat{s}_k + 1\}$. Likewise, $D_{\text{max}}(\hat{\mathcal{S}}^c_k)=\text{max}\{D^{(k)}_j: j<p-\hat{s}_k + 1\}$ is the largest adjacent distance in $\hat{\mathcal{S}}^c_k$. Then the final step of the SPSP can be implemented by finding these two sets such that the gap is greater than the largest adjacent distance in the irrelevant set ($\hat{\mathcal{S}}_k$), and smaller than the largest adjacent distance in the relevant set ($\hat{\mathcal{S}}^c_k$). In principle, the SPSP selection method adaptively finds a large enough gap to separate the relevant features from the irrelevant features. Specifically, it proposes the following criterion for identifying the relevant features $\frac{D_\text{max}(\hat{\mathcal{S}}_k)}{D(\hat{\mathcal{S}}_k, \hat{\mathcal{S}}^c_k)}\le R < \frac{D(\hat{\mathcal{S}}_k, \hat{\mathcal{S}}^c_k)}{D_{\text{max}}(\hat{\mathcal{S}}^c_k)}, $ where the constant control value $R$ can be estimated from the data. Finally, the SPSP method identifies a set of relevant variables $\tilde{\mathcal{S}}^{\text{SPSP}}$ as the
union of all $\hat{\mathcal{S}}_k$ for the tuning parameters $\lambda_1 < \lambda_2 < \cdots < \lambda_K$, i.e $\tilde{\mathcal{S}}^{\text{SPSP}}=\cup^K_{k=1} \hat{\mathcal{S}}_k, \label{eq:SPSP} $ as the estimate of the true model $\mathcal{S}_{0}$. Afterward, we refit the model to obtain the least square estimation $\hat{\boldsymbol{\beta}}^{\text{LS}}_{\tilde{\mathcal{S}}^{\text{SPSP}}}$ for the selected covariates $\tilde{\mathcal{S}}^{\text{SPSP}}$ but keep zero coefficients for unselected covariates. The estimation of coefficients for SPSP are $\hat{\boldsymbol{\beta}}^{\text{SPSP}}=(\hat{\boldsymbol{\beta}}^{\text{LS}}_{\tilde{\mathcal{S}}^{\text{SPSP}}},\boldsymbol{0}_{\hat{\mathcal{S}}^c}). \label{eq:SPSP_Beta} $

\section{Simulation Studies}\label{sec:simulation}

We conduct comprehensive simultation studies to investigate the performance of our SSCI in high-dimensional settings with independent or correlated covariates and low-dimensional settings with weak signals. Under each setting, we construct SSCIs using eight variable selection methods: SPSP based on the solution paths of SCAD (SPSP+SCAD) and MCP (SPSP+MCP), SCAD with 10-fold cross-validation (SCAD+CV), MCP with 10-fold cross-validation (MCP+CV), SPSP based on the solution paths of adaptive Lasso (SPSP+AdaLasso) and Lasso (SPSP+Lasso), adaptive Lasso with 10-fold cross-validation (AdaLasso+CV), and Lasso with 10-fold cross-validation (Lasso+CV). We also construct SCI debiased Lasso. We set the confidence level as 95\% for all the methods. Lastly, we construct a reference method assuming the true model is known. Although it is not available in practice, it can be used a benchmark for estimation uncertainty (denoted as ``Oracle''). With the known true model, the ``Oracle'' simultaneous confidence intervals use the Bonferroni adjustment to control the family-wise confidence level at 95\% for all the true signals. To obtain these results, we develop \texttt{SSCI} package to construct and visualize the SSCI based on the \texttt{glmnet} and \texttt{ggplot2} packages \citep{hastie_glmnet_2014, wickham_ggplot2_2011}. 
Below is a list of settings.

	\textbf{Study 1}: We generate 1000 data sets from the linear model with $\mathbf{X}_{j} \sim N(0, 1), \varepsilon_i \sim N(0, 1)$, for $i = 1,\dots,n$ and $j = 1,\dots, p$, and set $B=1000$. 
	\begin{itemize}\setlength{\itemindent}{1in}
		\setlength\itemsep{-0.5em}
		\item[\bf Example 1:]  (Independent covariates)  Let $n = 200$, $p = 300$, and $\boldsymbol{\beta}^{0}=(4, 3.5, 3, 2.5, 2, 0,...,0)$. 
		\item[\bf Example 2:]  (Correlated covariates) Let $n=50$, $p=100$, and $\boldsymbol{\beta}^{0}=(3, 2, 1.5,0,...,0)$. The pairwise covariate correlation is $\text{cor}(\mathbf{X}_j, \mathbf{X}_{j'})=0.5^{|j - {j'}|}$. 
		
		\item[\bf Example 3:] The setting is same as the Example 2 except the $n = 200$ and $p = 300$.
		
		\item[\bf Example 4:] (Correlated covariates with coefficients of alternating signs) Let $n = 200$, $p = 300$,  $\boldsymbol{\beta}^{0}=(0.9,-0.85,0.93,-1, 0.8, -0.85, 0.88, 0,...,0)$, $\text{cor}(\mathbf{X}_j, \mathbf{X}_{j'})=0.5^{|j - {j'}|}$.
	\end{itemize}

\begin{table}[H]
	\renewcommand{\arraystretch}{0.8} % this reduces the vertical spacing between rows
	\centering
	\caption{Simulation results of Examples 1 through 4}\label{table:sim1}
	\vspace{-5mm}
	\scalebox{0.62}{
			\begin{tabular}{@{}cl@{\extracolsep{5pt}} cccrcc}
				\multicolumn{8}{@{}p{20cm}@{}}{\footnotesize } \\
				\toprule
				\multicolumn{8}{c}{Study 1: High-dimensional settings} \\
				& & \multicolumn{3}{c}{Simultaneous Confidence Intervals} & \multicolumn{3}{c}{Model Confidence Bounds} \\
				\cmidrule{3-5}  \cmidrule{6-8}  
				Setting & Method &  $\text{P}^{\text{SCI}}_{\text{coverage}}(\%)$  & $\bar{w}_{\mathcal{S}_0}$ & $\bar{w}_{\mathcal{S}_0^c}$ & \multicolumn{1}{r}{$\text{P}^{\text{MCB}}_{\text{coverage}}(\%)$} & \multicolumn{2}{c}{$\bar{w}$} \\ 
				\midrule 
				& SSCI (SPSP+SCAD) & 93.7 & 0.353 (0.001) & {\color{blue} \textbf{0.000}} (0.000) & 100.0 & \multicolumn{2}{c}{0.00}   \\ 
				& SSCI (SPSP+MCP) & 93.8 & 0.353 (0.001) & {\color{blue} \textbf{0.000}} (0.000) & 100.0 & \multicolumn{2}{c}{0.00} \\ 
				\hfil Example 1 & SSCI (SPSP+AdaLasso) & 93.8 & 0.353 (0.001) & {\color{blue} \textbf{0.000}} (0.000) & 100.0 & \multicolumn{2}{c}{0.00}   \\ 
				\hfil $s$=5, $\rho$=0 & SSCI (SPSP+Lasso) & 93.7 & 0.353 (0.001) & {\color{blue} \textbf{0.000}} (0.000) & 100.0 & \multicolumn{2}{c}{0.00} \\ 
				\hfil $n$=200, $p$=300  & SSCI (SCAD+CV) & 97.6 & 0.454 (0.001) & 0.163 (0.003) & 100.0 & \multicolumn{2}{c}{148.56}   \\ 
				  & SSCI (MCP+CV) & 98.3 & 0.453 (0.001) & 0.162 (0.003) & 100.0 & \multicolumn{2}{c}{145.49}   \\ 
				& SSCI (AdaLasso+CV) & 94.1 & 0.354 (0.001) & 0.000 (0.000) & 100.0 & \multicolumn{2}{c}{0.00}   \\ 
				& SCI (hdi) & 0.0 & 0.283 (0.001) & 0.269 (0.001)  & --- &  \multicolumn{2}{c}{---}  \\ 
				& SCI (hdm) & 80.6 & 0.574 (0.002) & 0.513 (0.001)  & --- &  \multicolumn{2}{c}{---}  \\ 
				& SCI (Debiased Lasso) & 96.7 & 0.633 (0.001) & 0.633 (0.001)  & --- &  \multicolumn{2}{c}{---}  \\ 
				& Oracle & 95.3 & 0.370 (0.001) & 0.000 (0.000) & --- & \multicolumn{2}{c}{---}  \\ 	
				\cmidrule{1-8}
				& SSCI (SPSP+SCAD) & 92.6 & 0.870 (0.009) & {\color{blue} \textbf{0.003}} (0.001) & 99.5 & \multicolumn{2}{c}{0.59}   \\ 
				& SSCI (SPSP+MCP) & 91.9 & 0.829 (0.007) & {\color{blue} \textbf{0.001}} (0.001) & 99.9 & \multicolumn{2}{c}{0.26} \\ 
				\hfil Example 2 & SSCI (SPSP+AdaLasso) & 92.2 & 0.993 (0.012) & {\color{blue} \textbf{0.004}} (0.002) & 98.4 & \multicolumn{2}{c}{0.64}   \\ 
				\hfil $s$=3, $\rho$=0.5 & SSCI (SPSP+Lasso) & 92.1 & 0.991 (0.012) & {\color{blue} \textbf{0.004}} (0.002) & 98.2 & \multicolumn{2}{c}{0.66} \\ 
				\hfil $n$=50, $p$=100  & SSCI (SCAD+CV) & 97.1 & 1.117 (0.009) & 0.632 (0.005) & 99.9 & \multicolumn{2}{c}{93.31}   \\ 
				  & SSCI (MCP+CV) & 97.7 & 1.104 (0.008) & 0.664 (0.005) & 99.8 & \multicolumn{2}{c}{91.92}   \\ 
				& SSCI (AdaLasso+CV) & 95.3 & 0.931 (0.005) & 0.187 (0.004) & 100.0 & \multicolumn{2}{c}{38.79}   \\ 
				& SCI (hdi) & 0.8 & 0.714 (0.004) & 0.571 (0.003)  & --- &  \multicolumn{2}{c}{---}  \\ 
				& SCI (hdm) & 45.8 & 1.340 (0.010) & 1.119 (0.005)  & --- &  \multicolumn{2}{c}{---}  \\ 
				& SCI (Debiased Lasso) & 96.3 & 1.230 (0.005) & 1.230 (0.005)  & --- &  \multicolumn{2}{c}{---}  \\ 
				& Oracle & 94.2 & 0.851 (0.004) & 0.000 (0.000) & --- & \multicolumn{2}{c}{---}  \\ 	
				\cmidrule{1-8}
				& SSCI (SPSP+SCAD) & 94.2 & 0.394 (0.001) & {\color{blue} \textbf{0.000}} (0.000) & 100.0 & \multicolumn{2}{c}{0.00}   \\ 
				& SSCI (SPSP+MCP) & 94.1 & 0.395 (0.001) & {\color{blue} \textbf{0.000}} (0.000) & 100.0 & \multicolumn{2}{c}{0.00} \\ 
				\hfil Example 3 & SSCI (SPSP+AdaLasso) & 94.0 & 0.398 (0.002) & {\color{blue} \textbf{0.000}} (0.000) & 100.0 & \multicolumn{2}{c}{0.00}   \\ 
				\hfil $s$=3, $\rho$=0.5 & SSCI (SPSP+Lasso) & 93.9 & 0.398 (0.002) & {\color{blue} \textbf{0.000}} (0.000) & 100.0 & \multicolumn{2}{c}{0.00} \\ 
				\hfil $n$=200, $p$=300  & SSCI (SCAD+CV) & 98.9 & 0.543 (0.002) & 0.096 (0.002) & 100.0 & \multicolumn{2}{c}{92.54}   \\ 
				  & SSCI (MCP+CV) & 98.9 & 0.543 (0.002) & 0.095 (0.002) & 100.0 & \multicolumn{2}{c}{90.45}   \\ 
				& SSCI (AdaLasso+CV) & 93.6 & 0.393 (0.001) & 0.000 (0.000) & 100.0 & \multicolumn{2}{c}{0.00}   \\ 
				& SCI (hdi) & 0.0 & 0.344 (0.001) & 0.313 (0.001)  & --- &  \multicolumn{2}{c}{---}  \\ 
				& SCI (hdm) & 86.6 & 0.656 (0.002) & 0.672 (0.001)  & --- &  \multicolumn{2}{c}{---}  \\ 
				& SCI (Debiased Lasso) & 96.4 & 0.687 (0.001) & 0.687 (0.001)  & --- &  \multicolumn{2}{c}{---}  \\ 
				& Oracle & 95.3 & 0.411 (0.001) & 0.000 (0.000) & --- & \multicolumn{2}{c}{---}  \\ 	
				\cmidrule{1-8}
				& SSCI (SPSP+SCAD) & 95.7 & 0.614 (0.004) & {\color{blue} \textbf{0.296}} (0.004) & 100.0 & \multicolumn{2}{c}{256.63}   \\ 
				& SSCI (SPSP+MCP) & 95.7 & 0.615 (0.005) & {\color{blue} \textbf{0.334}} (0.005) & 100.0 & \multicolumn{2}{c}{255.21} \\ 
				\hfil Example 4 & SSCI (SPSP+AdaLasso) & 91.3 & 0.920 (0.008) & {\color{blue} \textbf{0.198}} (0.003) & 99.9 & \multicolumn{2}{c}{228.18}   \\ 
				\hfil $s$=7, $\rho$=0.5 & SSCI (SPSP+Lasso) & 91.1 & 0.922 (0.008) & {\color{blue} \textbf{0.197}} (0.003) & 99.9 & \multicolumn{2}{c}{227.89} \\ 
				\hfil $n$=200, $p$=300  & SSCI (SCAD+CV) & 80.7 & 0.778 (0.008) & 0.390 (0.001) & 99.6 & \multicolumn{2}{c}{293.80}   \\ 
				  & SSCI (MCP+CV) & 97.4 & 0.592 (0.003) & 0.381 (0.001) & 100.0 & \multicolumn{2}{c}{287.06}   \\ 
				& SSCI (AdaLasso+CV) & 88.8 & 0.603 (0.003) & 0.366 (0.001) & 99.2 & \multicolumn{2}{c}{293.23}   \\ 
				& SCI (hdi) & 0.0 & 0.345 (0.001) & 0.293 (0.001)  & --- &  \multicolumn{2}{c}{---}  \\ 
				& SCI (hdm) & 87.2 & 0.950 (0.004) & 1.007 (0.004)  & --- &  \multicolumn{2}{c}{---}  \\ 
				& SCI (Debiased Lasso) & 96.2 & 0.970 (0.002) & 0.970 (0.002)  & --- &  \multicolumn{2}{c}{---}  \\ 
				& Oracle & 94.2 & 0.486 (0.001) & 0.000 (0.000) & --- & \multicolumn{2}{c}{---}  \\ 	
				\bottomrule
				\multicolumn{8}{@{}p{23cm}@{}}{\small Note: SSCI is our Sparsified Simultaneous Confidence Intervals. We construct the SSCI using seven variable selection methods: SPSP based on solution paths of SCAD (SPSP+SCAD), MCP (SPSP+MCP), adaptive Lasso (SPSP+AdaLasso) and Lasso (SPSP+Lasso), SCAD with 10-fold cross-validation (SCAD+CV), MCP with 10-fold cross-validation (MCP+CV), adaptive Lasso with 10-fold cross-validation (AdaLasso+CV). For the alternative methods, we compare the simultaneous confidence intervals based on the de-sparsified Lasso method (SCI (hdi)), the SCI based on a double-selection approach of the Lasso method (SCI (hdm)), and the SCI based on the debiased Lasso method (SCI (Debiased Lasso)). We also compare SSCI with two reference methods. ``Oracle'' is simultaneous confidence intervals via the Bonferroni adjustment under the true model.} \\
			\end{tabular}
		}
\end{table}

We compare different methods in the following aspects: coverage probability of the simultaneous confidence intervals $\text{P}^{\text{SCI}}_{\text{coverage}}$, average interval width of true signals $\bar{w}_{\mathcal{S}_{0}}$ and non-signals $\bar{w}_{\mathcal{S}_{0}^c}$, MCB coverage probability $\text{P}^{\text{MCB}}_{\text{coverage}}$, average MCB width $\bar{w}$, and the coverage rate of the individual confidence interval for the weak signal $\text{P}^{\theta}_{\text{coverage}}$. We use the blue-boldfaced number to highlight the sparse or narrow intervals of the non-signals $\bar{w}_{\mathcal{S}_{0}^c}$. Results are shown in Table \ref{table:sim1} with the standard errors reported in parentheses.

In Table \ref{table:sim1}, all the SSCIs maintain valid coverage probabilities under high-dimensional settings. 
Their interval widths, on average, are narrower than all the alternative SCIs, including the best-performing SCI with debiased Lasso. 
In particular, the interval widths of the non-signals are close to zero (the blue-boldfaced numbers), especially in the first three examples. 
The zero-width intervals of the non-signals reveal that these intervals are sparse in the SSCI. 
These sparsified intervals provide an inference of the true model by implying the irrelevance of these covariates. 
Among all the penalized selection and non-convex penalized methods, the adoption of the SPSP selection method helps the SSCI approach achieve the best finite sample performance due to its advantages of stable selection and low false positive rate. 
In terms of the inference of the true model, the MCBs induced by the SSCI based on SPSP all maintain valid coverage probabilities. 
Their widths are also smaller than the rest. The zero-width of MCBs $\bar{w}= |\bar{\hat{\mathcal{S}}}_{2}|$ implies the upper bound model is same as the lower bound model, which is also the same as the true model. Our SSCIs can still perform well under scenarios with a high $p/n$ ratio or a larger number of non-zero regression coefficients. We put the additional simulation studies of these scenarios in our supplementary materials. 

Our approach can also work well when the weak signal covariates are involved. We follow \citet{shi_weak_2017} and include a study to show the advantages of our SSCIs for the weak signal situation.
{\bf Study 2}: We simulate 1000 data sets from linear model with $n = 100$ and $p = 20$. Let $\mathbf{X}_{j} \sim N(0, 1)$ and $\varepsilon_i \sim N(0, 2^2)$ for $i = 1,\dots,n$ and $j = 1,\dots, p$. Let $\boldsymbol{\beta}^{0}=(1,1,0.5,\theta,0,...,0)$ with a weak signal of $\theta=0.3$. The pairwise covariate correlation is $\text{cor}(\mathbf{X}_j, \mathbf{X}_{j'})=\rho^{|j - {j'}|}$. Results are shown in Table \ref{table:sim2}. 
	\begin{itemize}\setlength{\itemindent}{1in}
		\setlength\itemsep{-0.5em}
		\item[\bf Example 5:] (Independent covariates) $\rho=0$.
		\item[\bf Example 6:] (Weakly correlated covariates) $\rho=0.2$.
		\item[\bf Example 7:] (Moderately correlated covariates) $\rho=0.5$.
	\end{itemize}

\begin{table}[H]
	\renewcommand{\arraystretch}{0.8} % this reduces the vertical spacing between rows
	\centering
	\caption{Simulation results of Examples 5 through 7}\label{table:sim2}
	\scalebox{0.65}{
			\begin{tabular}{@{}cl@{\extracolsep{5pt}} cccrcc}
				\multicolumn{8}{@{}p{20cm}@{}}{\footnotesize } \\
				\toprule
				\multicolumn{8}{c}{Study 2: Weak signal settings} \\
				& & \multicolumn{3}{c}{Simultaneous Confidence Intervals} & \multicolumn{2}{c}{Model Confidence Bounds} & \multicolumn{1}{c}{Weak Signal} \\
				\cmidrule{3-5}  \cmidrule{6-7}   \cmidrule{8-8}  
				Setting & Method & $\text{P}^{\text{SCI}}_{\text{coverage}}(\%)$ & $\bar{w}_{\mathcal{S}_0}$ & $\bar{w}_{\mathcal{S}_0^c}$ & \multicolumn{1}{c}{$\text{P}^{\text{MCB}}_{\text{coverage}}(\%)$} &  \multicolumn{1}{c}{$\bar{w}$} & \multicolumn{1}{c}{$\text{P}^{\theta}_{\text{coverage}}(\%)$}  \\ 
				\cmidrule{1-8}
    % Example 5-1, N=1000; r=1000
				\hfil Example 5 & SSCI (SPSP+SCAD) & 96.0 &  1.428 (0.012) &  1.192 (0.005) & 100.0 & 19.23 & \multicolumn{1}{c}{99.8} \\ 
				\hfil $\rho$=0  & SSCI (SPSP+MCP) & 96.3 &  1.431 (0.012) & 1.197 (0.005) & 100.0 & 19.22 & \multicolumn{1}{c}{100.0} \\ 
				  & SSCI (SPSP+AdaLasso) & 88.5 & 1.459 (0.016) & 1.120 (0.007) & 99.8 & 19.86 & \multicolumn{1}{c}{99.8} \\ 
				& SSCI (SPSP+Lasso) & 78.8 & 1.347 (0.016) & 1.038 (0.007) & 99.8 & 19.74 & \multicolumn{1}{c}{97.5} \\ 
				& SSCI (SCAD+CV) & 98.3 & 1.369 (0.009) & 1.188 (0.004) & 99.8 & 18.54 & \multicolumn{1}{c}{99.8} \\ 
				& SSCI (MCP+CV) & 98.3 & 1.377 (0.009) & 1.195 (0.004) & 99.8 & 18.56 & \multicolumn{1}{c}{100.0} \\ 			
                & SSCI (AdaLasso+CV) & 94.0 & 1.335 (0.013) & 1.010 (0.004) & 100.0 & 19.20 & \multicolumn{1}{c}{100.0} \\ 
				& SCI (hdi) & 39.0 & 0.828 (0.004) & 0.817 (0.004)  & --- &  --- & \multicolumn{1}{c}{94.5} \\ 
			    & SCI (hdm) & 87.0 & 1.204 (0.008) & 1.201 (0.005)  & --- &  --- & \multicolumn{1}{c}{93.7} \\ 
				\hfil & SCI (Debiased Lasso) & 96.3 & 1.372 (0.005) & 1.372 (0.005) &  ---  &  ---  & \multicolumn{1}{c}{99.5} \\ 
				& Oracle & 93.5 & 1.026 (0.004) & 0.000 (0.000) &  ---  &  ---  & \multicolumn{1}{c}{98.0} \\ 
                \cmidrule{1-8}
% Example 6-2, N=1000; r=1000
				\hfil Example 6 & SSCI (SPSP+SCAD) & 94.2 & 1.372 (0.008) & 1.155 (0.003) & 100.0 & 19.13 & \multicolumn{1}{c}{99.6} \\ 
				\hfil $\rho$=0.2 & SSCI (SPSP+MCP) & 95.3 &  1.376 (0.008) & 1.161 (0.003) & 100.0 & 19.16 & \multicolumn{1}{c}{99.8} \\ 
				  & SSCI (SPSP+AdaLasso) & 92.1 & 1.413 (0.009) & 1.118 (0.004) & 100.0 & 19.56 & \multicolumn{1}{c}{97.9} \\
				& SSCI (SPSP+Lasso) & 84.6 & 1.405 (0.011) & 1.024 (0.005) & 99.8 & 19.65 & \multicolumn{1}{c}{98.1} \\ 
				& SSCI (SCAD+CV) & 96.1 & 1.295 (0.006) & 1.145 (0.003) & 99.4 & 18.61 & \multicolumn{1}{c}{99.8} \\ 
				& SSCI (MCP+CV) & 96.0 & 1.303 (0.006) & 1.158 (0.003) & 99.6 & 18.66 & \multicolumn{1}{c}{99.5} \\ 
			    & SSCI (AdaLasso+CV) & 93.5 & 1.234 (0.008) & 0.947 (0.003) & 99.9 & 18.82 & \multicolumn{1}{c}{100.0} \\ 
				& SCI (hdi) & 34.2 & 0.827 (0.002) & 0.812 (0.002)  & --- &  --- & \multicolumn{1}{c}{93.7} \\ 
				& SCI (hdm) & 87.5 & 1.211 (0.004) & 1.225 (0.002)  & --- &  --- & \multicolumn{1}{c}{99.6} \\ 
				\hfil & SCI (Debiased Lasso) & 96.1 & 1.417 (0.003) & 1.417 (0.003) &  ---  &  ---  & \multicolumn{1}{c}{99.8} \\ 
				& Oracle & 95.1 & 1.056 (0.003) & 0.000 (0.000) &  ---  &  ---  & \multicolumn{1}{c}{98.7} \\ 
				\cmidrule{1-8}
    
% Example 7-2, N=1000; r=1000
				\hfil Example 7 & SSCI (SPSP+SCAD) & 95.3 &  1.569 (0.007) & 1.445 (0.004) & 99.4 & 18.81 & \multicolumn{1}{c}{99.6} \\ 
				\hfil $\rho$=0.5 & SSCI (SPSP+MCP) & 93.3 &  1.598 (0.008) & 1.456 (0.004) & 99.4 & 18.95 & \multicolumn{1}{c}{99.4} \\ 
				  & SSCI (SPSP+AdaLasso) & 91.7 & 1.741 (0.012) & 1.319 (0.004) & 100.0 & 19.72 & \multicolumn{1}{c}{99.9} \\
				& SSCI (SPSP+Lasso) & 95.0 & 1.586 (0.029) & 0.915 (0.024) & 100.0 & 19.69 & \multicolumn{1}{c}{100.0} \\ 
				& SSCI (SCAD+CV) & 91.5 & 1.625 (0.010) & 1.351 (0.004) & 99.8 & 19.27 & \multicolumn{1}{c}{99.5} \\ 
				& SSCI (MCP+CV) & 89.7 & 1.636 (0.010) & 1.373 (0.004) & 99.6 & 19.34 & \multicolumn{1}{c}{99.7} \\ 
			    & SSCI (AdaLasso+CV) & 97.9 & 1.521 (0.008) & 1.300 (0.003) & 99.9 & 19.04 & \multicolumn{1}{c}{99.7}  \\ 
				& SCI (hdi) & 34.4 & 0.950 (0.003) & 0.926 (0.002)  & --- &  --- & \multicolumn{1}{c}{93.2} \\ 
				& SCI (hdm) & 87.7 & 1.463 (0.007) & 1.493 (0.004)  & --- &  --- & \multicolumn{1}{c}{99.2} \\ 
				\hfil   & SCI (Debiased Lasso) & 94.9 & 1.749 (0.004) & 1.749 (0.004) &  ---  &  ---  & \multicolumn{1}{c}{99.7}  \\ 
				& Oracle & 94.6 & 1.253 (0.004) & 0.000 (0.000) &  ---  &  ---  & \multicolumn{1}{c}{98.4} \\ 
				\bottomrule
				
				\multicolumn{8}{@{}p{23cm}@{}}{\small Note: SSCI is our Sparsified Simultaneous Confidence Intervals. We construct the SSCI using seven variable selection methods: SPSP based on solution paths of SCAD (SPSP+SCAD), MCP (SPSP+MCP), adaptive Lasso (SPSP+AdaLasso) and Lasso (SPSP+Lasso), SCAD with 10-fold cross-validation (SCAD+CV), MCP with 10-fold cross-validation (MCP+CV), adaptive Lasso with 10-fold cross-validation (AdaLasso+CV). For the alternative methods, we compare the simultaneous confidence intervals based on the de-sparsified Lasso method (SCI (hdi)), the SCI based on a double-selection approach of the Lasso method (SCI (hdm)), and the SCI based on the debiased Lasso method (SCI (Debiased Lasso)). We also compare SSCI with two reference methods. ``Oracle'' is simultaneous confidence intervals via the Bonferroni adjustment under the true model.} \\
			\end{tabular}
		}
\end{table}

Under weak signal settings, the advantages of our SPSP-based SSCIs persist. Although the interval widths are inflated compared to Study 1, our SSCIs are still narrower than the best alternative SCIs, the SCI (Debiased Lasso), for both the true and non-signals. In Table 3, most of our SSCIs have good coverage rates and the average width of intervals of non-signals is noticeably narrower than that of the SCI (Debiased Lasso). Besides, the MCBs are still tight and can achieve nominal coverage probability. Lastly, the individual confidence interval for the weak signal all have valid coverage probabilities. 

\section{Real Data Examples}\label{sec:real-data-examples}

In this section, we apply the proposed approach to investigate biology's critical genome-wide transcriptional regulation problem. Specifically, biologists are interested in identifying a few crucial transcription factors (TFs) that are associated with the gene expression levels during the yeast cell cycle process. The response variable of this data is the $n=1132$ gene expression levels of yeast in the study \citet{spellman_comprehensive_1998} and \citet{ luan_clustering_2003}. The covariates include 96  transcription factors (TFs) measured by binding probabilities using a mixture model based on the ChiP data \citep{lee_transcriptional_2002, wang_group_2007} (standardized to have zero mean and unit variance) and their interaction effects. The dataset is publicly available in the R package \texttt{PGEE} \citep{inan_pgee_2017}. 

Previous studies have focused on either the individual TF effects or the synergistic effects where a pair of TFs cooperate to regulate transcription in the cycle process
\citep{wang_group_2007, cheng_systematic_2008, wang_penalized_2012, das_interacting_2004}. For example, \citet{banerjee_identifying_2003} and \citet{tsai_statistical_2005} identify 31 and 18 cooperative TF pairs, respectively. However, to the best of our knowledge, there is no simultaneous inference of both the individual TFs and cooperative TF pairs. 

We attempt to conduct simultaneous inference to investigate this issue. In particular, we pre-screen all the individual TFs and 4560=96*95/2 TF interactions using sure independence screening \citep{fan_sure_2008} and identify 1200 covariates correlated with the response variable. We then add the individual TF back even if the screening does not select them to avoid missing any vital TFs. We obtain 1263 covariates, including the time $t$, for our inference. We construct SSCI using the recommended SPSP + AdaLasso with $B=5000$.

\begin{figure}[t]
		\centering \includegraphics[width=0.8\textwidth]{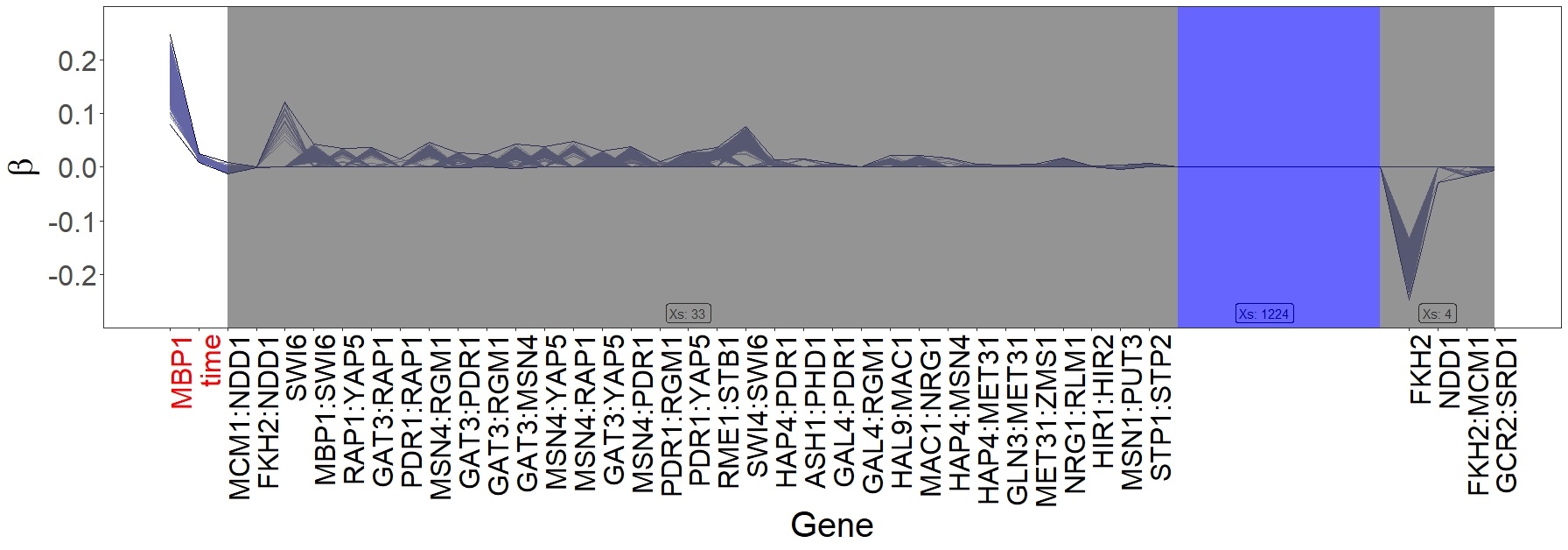} 
		\caption{Graphical presentation of the SSCI (SPSP+AdaLasso) for the yeast cell-cycle data.} \label{fig:YeastG1_Sparsified_SCI}
\end{figure}

\begin{table}[t]
\normalsize
\centering
\caption{Real data example: Yeast cell-cycle (G1) data}\label{table:application}
\scalebox{0.45}{
\begin{tabular}{cccccccccccccc}
\toprule
& \multicolumn{13}{c}{\Large $\text{SSCI}_{0.95}$}  \\
\cmidrule{2-14}  
\Large $\text{SSCI}_{0.95}$ (SPSP+AdaLasso) & \multicolumn{4}{c}{\Large Confidence intervals} & \multicolumn{5}{c}{\Large Literature evidence} & \multicolumn{4}{c}{\Large Effects documented in the literature} \\
\midrule	
\Large time        & \multicolumn{4}{c}{ \Large Significantly +++ } & \multicolumn{5}{c}{ \Large } & \multicolumn{4}{c}{ \Large } \\
\Large MBP1        & \multicolumn{4}{c}{ \Large Significantly +++ } & \multicolumn{5}{c}{ \Large \citet{tsai_statistical_2005}, \citet{wang_group_2007}, \citet{yue_two-step_2019}} & \multicolumn{4}{c}{ \Large  + }  \\
\Large SWI6        & \multicolumn{4}{c}{ \Large Plausibly +} & \multicolumn{5}{c}{ \Large \citet{wang_group_2007}, \citet{yue_two-step_2019}} & \multicolumn{4}{c}{ \Large  + } \\
\Large FKH2        & \multicolumn{4}{c}{ \Large Plausibly -} & \multicolumn{5}{c}{ \Large \citet{tsai_statistical_2005}, \citet{wang_group_2007}, \citet{yue_two-step_2019}} & \multicolumn{4}{c}{ \Large  - } \\
\Large NDD1        & \multicolumn{4}{c}{ \Large Plausibly -} & \multicolumn{5}{c}{ \Large \citet{tsai_statistical_2005}, \citet{wang_group_2007}, \citet{yue_two-step_2019}} & \multicolumn{4}{c}{ \Large  - } \\ 
\Large MBP1-SWI6   & \multicolumn{4}{c}{ \Large Plausibly +} & \multicolumn{5}{c}{ \Large \citet{koch_role_1993}, \citet{banerjee_identifying_2003}, \citet{tsai_statistical_2005}} & \multicolumn{4}{c}{ \Large  + } \\ 
\Large MSN4-YAP5   & \multicolumn{4}{c}{ \Large Plausibly +} & \multicolumn{5}{c}{ \Large \citet{banerjee_identifying_2003}, \citet{tsai_statistical_2005}} & \multicolumn{4}{c}{ \Large +} \\
\Large MSN4-PDR1   & \multicolumn{4}{c}{ \Large Plausibly +} & \multicolumn{5}{c}{ \Large \citet{tsai_statistical_2005}} & \multicolumn{4}{c}{ \Large +} \\
\Large SWI4-SWI6   & \multicolumn{4}{c}{ \Large Plausibly +} & \multicolumn{5}{c}{ \Large \citet{koch_role_1993}, \citet{banerjee_identifying_2003}, \citet{tsai_statistical_2005}} & \multicolumn{4}{c}{ \Large +} \\
\Large HAP4-PDR1   & \multicolumn{4}{c}{ \Large Plausibly +} & \multicolumn{5}{c}{ \Large \citet{tsai_statistical_2005}} & \multicolumn{4}{c}{ \Large +} \\
\Large HIR1-HIR2   & \multicolumn{4}{c}{ \Large Plausibly +} & \multicolumn{5}{c}{ \Large \citet{loy_ndd1_1999}, \citet{banerjee_identifying_2003}} & \multicolumn{4}{c}{ \Large NA} \\
\Large FKH2-MCM1   & \multicolumn{4}{c}{ \Large Plausibly -} & \multicolumn{5}{c}{ \Large \citet{kumar_forkhead_2000}, \citet{banerjee_identifying_2003}} & \multicolumn{4}{c}{ \Large NA} \\
\Large GAT3-PDR1   & \multicolumn{4}{c}{ \Large Plausibly} & \multicolumn{5}{c}{ \Large \citet{banerjee_identifying_2003}, \citet{tsai_statistical_2005}} & \multicolumn{4}{c}{ \Large +} \\ 
\Large GAT3-MSN4   & \multicolumn{4}{c}{ \Large Plausibly} & \multicolumn{5}{c}{ \Large \citet{banerjee_identifying_2003}} & \multicolumn{4}{c}{ \Large NA} \\
\Large MCM1-NDD1   & \multicolumn{4}{c}{ \Large Plausibly} & \multicolumn{5}{c}{ \Large \citet{kumar_forkhead_2000} \citet{banerjee_identifying_2003}, \citet{tsai_statistical_2005}} & \multicolumn{4}{c}{ \Large -} \\
\Large FKH2-NDD1   & \multicolumn{4}{c}{ \Large Plausibly} & \multicolumn{5}{c}{ \Large \citet{kumar_forkhead_2000}, \citet{banerjee_identifying_2003}, \citet{tsai_statistical_2005}} & \multicolumn{4}{c}{ \Large NA} \\ 

	\midrule
	& \multicolumn{13}{c}{\Large $\text{MCB}_{0.95}$}  \\
	\cmidrule{2-14}  
	& \multicolumn{6}{c}{\Large Lower confidence bound model} & \multicolumn{7}{c}{\Large Upper confidence bound model}  \\
	\cmidrule(r){2-7}  \cmidrule(l){8-14}  
	\Large SSCI (SPSP+AdaLasso) & \multicolumn{6}{c}{\Large \{\text{time, MBP1}\}} & \multicolumn{7}{c}{\Large \{\text{time, MBP1, and all other plausible covariates}\}}  \\
	\bottomrule
	\multicolumn{14}{@{}p{35cm}@{}}{\large Note: We provide the literature evidence for our selected cooperative TF pairs if either one or both TFs of the pair have been identified in the literature. We display ``Significantly positive/negative'' as ``Significantly +++/- - -'', and ``Plausibly positive/negative'' as ``Plausibly +/-''. The last column shows the sign of the coefficients documented in the corresponding literature.} \\
\end{tabular}
}
\end{table}

We present the results of 95\% SSCI in Figure \ref{fig:YeastG1_Sparsified_SCI}. The upper and lower bounds of the confidence intervals are given in Table 3 of the supplementary materials. Among the 1263 covariates, only two covariates, MBP1 and time, are identified as significant covariates, deemed important for regulating the transcription process. Another 37 covariates are identified as plausible covariates, whose relevance to the gene expression levels requires further studies. The rest of the covariates are identified as unimportant. All covariates are displayed in Figure \ref{fig:YeastG1_Sparsified_SCI} using different shaded areas. Our results decrease the size of the candidate pool of biologically relevant TFs and synergistic interactions. 

Comparing our results with the literature, all of our identified individual TFs, MBP1, FKH2, NDD1, and SWI6, are experimentally verified \citep{wang_group_2007}. Besides, 12 of our identified cooperative TF pairs have been documented in the literature (MBP1-SWI6, GAT3-PDR1, GAT3-MSN4, MSN4-YAP5, MSN4-PDR1, SWI4-SWI6, HAP4-PDR1, GAL4-RGM1, HIR1-HIR2, MCM1-NDD1, FKH2-NDD1, FKH2-MCM1). The remaining 22 plausible TF pairs have either one or both TFs selected in the literature \citep{wang_group_2007} except the PDR1-RAP1 pair. We present some representative covariates from our SSCI in comparison to the literature in Table \ref{table:application}. A detailed comparison of all plausible covariates is given in Table 4 of the supplementary material.

For these plausible covariates, one advantage of SSCI is that it reports the possible sign of their effects if one of their boundaries is at zero, e.g., [0,1] or [-1,0]. We report the estimated signs of TFs as ``plausibly +/-'' in Table \ref{table:application} and Table 4 of the supplementary materials and further compare the estimated signs with the documented regulatory effects in the literature. Most identified TFs and TF pairs (MBP1, FKH2, NDD1, SWI6, MBP1-SWI6, MSN4-YAP5, MSN4-PDR1, SWI4-SWI6, and HAP4-PDR1) are consistent with the literature \citep{tsai_statistical_2005}.

\section{Conclusion}
\label{sec:conc}

This article proposes a sparse version of the simultaneous confidence intervals for high-dimensional linear regression. 
The proposed confidence intervals reflect rich information about the parameter and the model. 
Our approach has been theoretically and empirically justified, with desirable asymptotic properties and satisfactory numerical performance. 
There are many potential directions for future research. 
First of all, the proposed procedure is not limited to the regression method. For example, the proposed method can be similarly applied to the Gaussian graphical models \citep{wang_CG_2021}. 
In such a setting, we could construct the sparsified simultaneous confidence intervals (SSCI) for all the elements of the sparse precision matrix, $\theta_{ij}$. 
By applying an appropriate bootstrap procedure and a consistent graphical model selection method, we believe the SSCI for the precision matrix will have similar functionalities as the proposed SSCI for regression.

Second, we observe the varying performance of inference based on different variable selection methods. 
It would be very interesting to examine how applying different selection methods to the same data set might yield different results. 
Meanwhile, the weak signals pose challenges to many selection methods.
Although we have tested our method numerically, extending our theoretical results to the weak signal case is an important topic \citep{liu_bootstrap_2020}.
In addition, our approach may be easily extended to build simultaneous confidence intervals for a subset of covariates, which is of great value in many real problems \citep{zhang_simultaneous_2017}.

\bigskip
\begin{center}
{\large\bf SUPPLEMENTARY MATERIAL}
\end{center}

\begin{description}

\item[Supplementary file:] The supplementary file (pdf) contains the algorithm of the residual bootstrap, details of two real data examples, and technical proofs. We demonstrate the SSCI on the low-dimensional Boston housing data. In addition, we include more details of the high-dimensional Yeast cell-cycle (G1) data analysis in the paper. We also present three additional simulation settings in this file. 

\item[R-package for sparsified simultaneous confidence intervals:] We develop the R-package \texttt{SSCI} to construct and visualize the sparsified simultaneous confidence intervals described in the article. This package implements the SSCI method proposed in this paper and supports building the SSCI using all selection approaches adopted in simulation studies. (R package binary file)

\end{description}

On behalf of all authors, the corresponding author states that there is no conflict of interest. 

\begingroup\footnotesize
\let\section\subsubsection
\makeatletter
\renewcommand\@openbib@code{\itemsep\z@}
\makeatother
\bibliographystyle{plainnat}

\bibliography{bibliography}
\endgroup

\end{document}

\def\spacingset#1{\renewcommand{\baselinestretch}%
{#1}\small\normalsize} \spacingset{1}

%%%%%%%%%%%%%%%%%%%%%%%%%%%%%%%%%%%%%%%%%%%%%%%%%%%%%%%%%%%%%%%%%%%%%%%%%%%%%%

\if0\blind
{
  \title{\bf Supplementary Materials for ``Sparsified Simultaneous Confidence Intervals for High-Dimensional Linear Models''}
  
  \author{Xiaorui Zhu, Yichen Qin, and Peng Wang\thanks{Xiaorui Zhu is an Assistant Professor in the Department of Business Analytics \& Technology Management, Towson University. Yichen Qin is a Professor in the Department of Operations, Business Analytics, and Information Systems at the University of Cincinnati. Peng Wang is an Associate Professor in the Department of Operations, Business Analytics, and Information Systems at the University of Cincinnati.} 
  %\\
  %University of Cincinnati
  }
  \maketitle
} \fi

\if1\blind
{
  \bigskip
  \bigskip
  \bigskip
  \begin{center}
    {\LARGE\bf Sparsified Simultaneous Confidence Intervals for High-Dimensional Linear Models}
\end{center}
  \medskip
} \fi

\bigskip

\newpage
\spacingset{1.5} % DON'T change the spacing!
\section{Residual Bootstrap Algorithm}
\label{sec:residual}
The residual bootstrap is part of our algorithm for constructing the sparsified simultaneous confidence intervals.
The details of residual bootstrap are outlined as follows.

\begin{algo}
\caption{Residual Bootstrap Method}\label{alg:residBoot}
\hline \\
\vspace*{-12pt}
\begin{tabbing}
   \qquad \enspace Step 1: Apply a two-stage estimation procedure on $\left\{\mathbf{y},\mathbf{X}\right\}$ to obtain \\ \qquad \qquad \qquad the selected model $\tilde{\mathcal{S}}$ and its refitted coefficient estimate $\tilde{\boldsymbol{\beta}}$; \\
   
   \qquad \enspace Step 2: Compute residuals: $\tilde{\boldsymbol{\varepsilon}} = \mathbf{y} - \mathbf{X}\tilde{\boldsymbol{\beta}}$; \\
   
   \qquad \enspace Step 3: Centralize residuals: $\tilde\varepsilon_{\text{cent},i} = \tilde\varepsilon_{i} - \bar{\tilde\varepsilon} \; (i=1,\dots,n)$,  where $\bar{\tilde\varepsilon}=n^{-1}\sum\tilde\varepsilon_i$;  \\
   
   \qquad \enspace Step 4: Resample B copies of $\tilde{\boldsymbol{\varepsilon}}^{(b)}=(\varepsilon^{(b)}_1, \dots,\varepsilon^{(b)}_{n})$  from $\tilde\varepsilon_{\text{cent},i}$,  where $b=1,\dots,B$; \\
   
   \qquad \enspace Step 5: Construct bootstrapped response as:  \\ \qquad \qquad \qquad $\mathbf{y}^{(b)} =\mathbf{X}\tilde{\boldsymbol{\beta}}+\tilde{\boldsymbol{\varepsilon}}^{(b)}$ to obtain bootstrap samples $\{(\mathbf{y}^{(b)},\mathbf{X}, \tilde{\boldsymbol{\varepsilon}}^{(b)})\}^B_{b=1}$; \\
   
   \qquad \enspace Step 6: Apply the same variable selection in a two-stage fashion on $b$-th bootstrap sample to \\ \qquad \qquad \qquad obtain bootstrap model $\hat{\mathcal{S}}^{(b)}$ and its refitted bootstrap estimate $\hat{\boldsymbol{\beta}}^{(b)}$.
\end{tabbing}
\vspace*{-5pt}
\hline \\
\end{algo}

\section{Real Data Example: Boston Housing Data}\label{s:real-data-examples}
We demonstrate the proposed method on the low-dimensional Boston housing data. The Boston housing data \citep{harrison_hedonic_1978} contains median values of owner-occupied homes in 506 U.S. census tracts in Boston and 13 covariates, including CRIM, ZN, INDUS, CHAS, NOX, RM, AGE, DIS, RAD, TAX, PTRATIO, BLACK, and LSTAT (see the supplementary materials for details). We remove observations with a crime rate over 3.2 as suggested in previous studies \citep{li1991sliced, chen_coordinate-independent_2010, qin_penalized_2017} and apply the proposed method to the remaining 374 observations. The results are shown in Table \ref{table:1}.

\begin{table}[!h]
\normalsize
\centering
\caption{Boston housing data}\label{table:1}
\scalebox{0.4}{
\begin{tabular}{cccccccccccccc}
	\toprule
    & \multicolumn{13}{c}{\Large $\text{SSCI}_{0.95}$}  \\
	\cmidrule{2-14}  
\Large Selection Method & \Large LSTAT & \Large ZN & \Large INDUS & \Large CHAS & \Large NOX & \Large RM & \Large AGE & \Large DIS & \Large RAD & \Large TAX & \Large PTRATIO & \Large BLACK & \Large CRIM \\ 
	\midrule
\Large SSCI (SPSP+AdaLasso) & \large [-0.28,0] & \large [0,0.06] & \large \{0\} & \large [0,0.04] & \large \{0\} & \large [0.58,0.94] & \large \{0\} & \large [0,0.05] & \large \{0\} & \large \{0\} & \large [-0.23,0] & \large \{0\} & \large \{0\} \\
\Large SSCI (SPSP+Lasso) & \large [-0.30,0] & \large [0,0.03] & \large \{0\} & \large \{0\} & \large \{0\} & \large [0.57,0.97] & \large \{0\} & \large \{0\} & \large \{0\} & \large \{0\} & \large [-0.24,0] & \large \{0\} & \large \{0\} \\
% \Large SSCI (AdaLasso+CV) & \large [-0.33,-0.12] & \large [-0.02,0.13] & \large [-0.05,0.07] & \large [-0.02,0.08] & \large [-0.25,0] & \large [0.51, 0.70] & \large [-0.15,0] & \large [-0.33,-0.09] & \large [0.02,0.14] & \large [-0.19,-0.07] & \large [-0.24,-0.11] & \large [-0.01,0.10] & \large [-0.01,0.17]  \\
% \Large SSCI (Lasso+CV) & \large  [-0.33,-0.11] & \large [-0.02,0.13] & \large [-0.02,0.08] & \large [-0.02,0.09] & \large [-0.26,-0.03] & \large [0.51, 0.68] & \large [-0.15,-0.01] & \large [-0.36,-0.14] & \large [0.02,0.14] & \large [-0.20,-0.08] & \large [-0.24,-0.11] & \large [-0.01,0.11] & \large [0,0.18]  \\
\Large SCI (Debiased Lasso) & \large [-0.32,-0.12] & \large [-0.02,0.14] & \large [-0.09,0.08] & \large [-0.02,0.09] & \large [-0.26,-0.03] & \large [0.50, 0.69] & \large [-0.16,-0.01] & \large [-0.37,-0.15] & \large [0.02,0.14] & \large [-0.20,-0.07] & \large [-0.24,-0.11] & \large [-0.01,0.11] & \large [-0.01,0.18]  \\
	\midrule
	& \multicolumn{13}{c}{\Large \text{MCB}_{0.95}}  \\
	\cmidrule{2-14}  
	& \multicolumn{6}{c}{\Large Lower confidence bound model} & \multicolumn{7}{c}{\Large Upper confidence bound model}  \\
	\cmidrule(r){2-7}  \cmidrule(l){8-14}  
	\Large SSCI (SPSP+AdaLasso) & \multicolumn{6}{c}{\Large $\{\text{RM}\}$} & \multicolumn{7}{c}{\Large $\{\text{RM, LSTAT, ZN, PIRATIO, CHAS, DIS}\}$}  \\
	\Large SSCI (SPSP+Lasso) & \multicolumn{6}{c}{\Large $\{\text{RM}\}$} & \multicolumn{7}{c}{\Large $\{\text{RM, LSTAT, ZN, PIRATIO}\}$}  \\
% 	\Large SSCI (AdaLasso+CV) & \multicolumn{6}{c}{\Large $\{\text{LSTAT, RM, DIS, RAD, TAX, PTRATIO}\}$} & \multicolumn{7}{c}{\Large $\{\text{All variables}\}$}  \\
% 	\Large SSCI (Lasso+CV) & \multicolumn{6}{c}{\Large $\{\text{LSTAT, NOX, RM, AGE, DIS, RAD, TAX, PTRATIO}\}$} & \multicolumn{7}{c}{\Large $\{\text{All variables}\}$}  \\
	\Large SCI (Debiased Lasso) & \multicolumn{6}{c}{\Large $\{\text{RM, LSTAT, NOX, AGE, DIS, RAD, TAX, PTRATIO}\}$} & \multicolumn{7}{c}{\Large $\{\text{All variables}\}$}  \\
	\bottomrule
\end{tabular}
}
\end{table}

\begin{figure}[!h]
		\centering \includegraphics[width=0.7\textwidth]{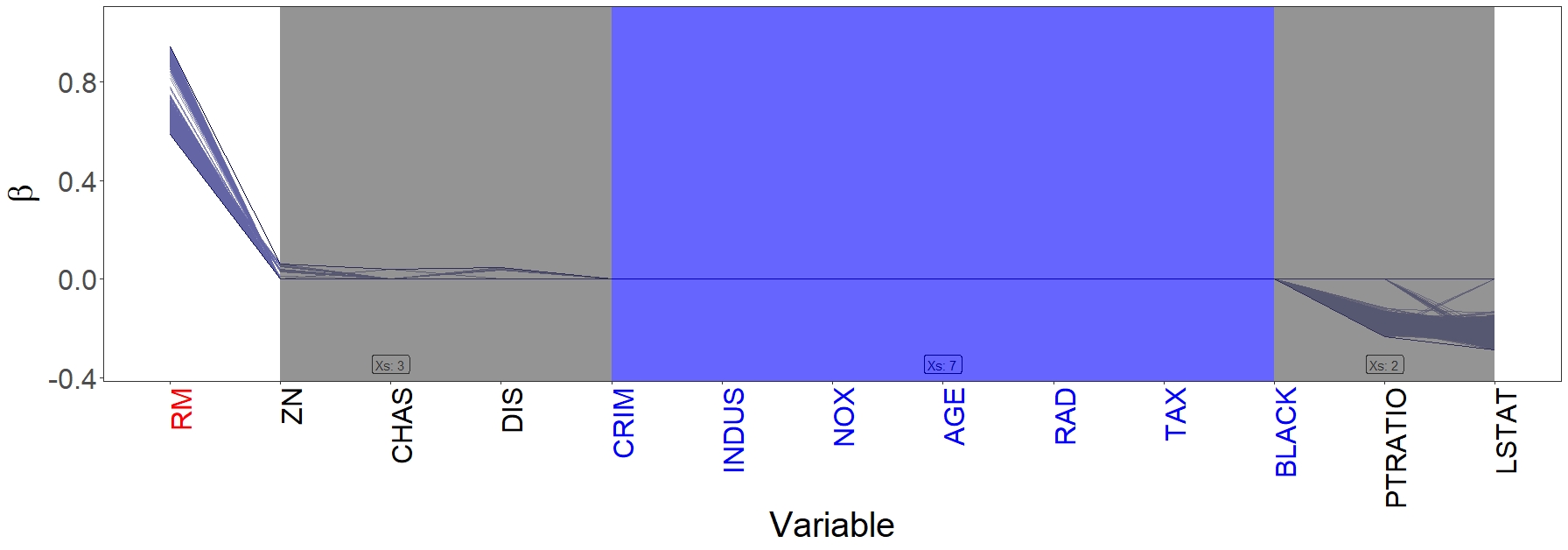} 
	\caption{Graphical presentation of the boston housing data.}\label{fig:Boston_Sparsified_SCI}
\end{figure}

Based on the SSCIs, we find interesting insights into the important factors related to housing prices. First, the significance of covariate RM implies that the average number of rooms is positively associated with the house's value. Second, the LSTAT, ZN, CHAS, DIS, and PTRATIO are plausible covariates that deserve further exploration. Third, the rest of the covariates should be excluded from the analysis since their confidence intervals are shrunken to zero, marked as $\{0\}$. The SSCI plot in Figure \ref{fig:Boston_Sparsified_SCI} provides the same conclusions. In contrast, all the confidence intervals of SCI debiased Lasso have non-zero width, which cannot identify the covariates to be excluded. 

\section{Real Data Example: Yeast Cell-Cycle (G1) Data }
\label{sec:real_data_example}

For the yeast cell-cycle data set, we present the upper and lower bounds of 95\% SSCI in Table \ref{table:application}. We use bold fonts to emphasize the bounds shrunken to zero. If the confidence interval contains zero, but the upper and lower bounds are not shrunken to zero, we use italic fonts.

	\begin{table}[H]
		\centering
		\caption{SSCI of Yeast cell-cycle (G1) data}\label{table:application}
		\scalebox{0.7}{
			\begin{tabular}{cccccccccc}
				\toprule
				Covariates & \multicolumn{4}{c}{Confidence intervals} & Covariates & \multicolumn{4}{c}{Confidence intervals}  \\ \hline
				time        & \multicolumn{4}{c}{[\bf 0.0094, 0.0244]} \\
				MBP1        & \multicolumn{4}{c}{[\bf 0.0789, 0.2476]} \\ 
				SWI6        & \multicolumn{4}{c}{[{\bf 0.0000}, 0.1207]} & GAL4-PDR1   & \multicolumn{4}{c}{[{\bf  0.0000}, 0.0068]} \\ 
				FKH2        & \multicolumn{4}{c}{[-0.2475, {\bf 0.0000}]} & GAL4-RGM1   & \multicolumn{4}{c}{[{\bf  0.0000}, 0.0014]}  \\
				NDD1        & \multicolumn{4}{c}{[-0.0290, {\bf 0.0000}]} & HAL9-MAC1   & \multicolumn{4}{c}{[{\bf  0.0000}, 0.0219]} \\
				MBP1-SWI6   & \multicolumn{4}{c}{[{\bf  0.0000}, 0.0433]} & MAC1-NRG1   & \multicolumn{4}{c}{[{\bf  0.0000}, 0.0215]}  \\
				RAP1-YAP5   & \multicolumn{4}{c}{[{\bf  0.0000}, 0.0348]} & HAP4-MSN4   & \multicolumn{4}{c}{[{\bf  0.0000}, 0.0167]} \\
				GAT3-RAP1   & \multicolumn{4}{c}{[{\bf  0.0000}, 0.0361]} & HAP4-MET31  & \multicolumn{4}{c}{[{\bf  0.0000}, 0.0054]} \\
				PDR1-RAP1   & \multicolumn{4}{c}{[{\bf  0.0000}, 0.0160]} & GLN3-MET31  & \multicolumn{4}{c}{[{\bf  0.0000}, 0.0041]} \\
				MSN4-RGM1   & \multicolumn{4}{c}{[{\bf  0.0000}, 0.0458]} & MET31-ZMS1  & \multicolumn{4}{c}{[{\bf  0.0000}, 0.0055]} \\
				GAT3-RGM1   & \multicolumn{4}{c}{[{\bf  0.0000}, 0.0240]} & NRG1-RLM1   & \multicolumn{4}{c}{[{\bf  0.0000}, 0.0175]} \\
				MSN4-YAP5   & \multicolumn{4}{c}{[{\bf  0.0000}, 0.0382]} & HIR1-HIR2   & \multicolumn{4}{c}{[{\bf  0.0000}, 0.0022]} \\
				MSN4-RAP1   & \multicolumn{4}{c}{[{\bf  0.0000}, 0.0484]} & STP1-STP2   & \multicolumn{4}{c}{[{\bf  0.0000}, 0.0068]} \\
				GAT3-YAP5   & \multicolumn{4}{c}{[{\bf  0.0000}, 0.0305]} & FKH2-MCM1   & \multicolumn{4}{c}{[-0.0169, {\bf  0.0000}]} \\
				MSN4-PDR1   & \multicolumn{4}{c}{[{\bf  0.0000}, 0.0382]} & GCR2-SRD1   & \multicolumn{4}{c}{[-0.0059, {\bf  0.0000}]} \\
				PDR1-RGM1   & \multicolumn{4}{c}{[{\bf  0.0000}, 0.0109]} & GAT3-PDR1   & \multicolumn{4}{c}{[{\it -0.0010, 0.0272}]} \\
				PDR1-YAP5   & \multicolumn{4}{c}{[{\bf  0.0000}, 0.0281]} & GAT3-MSN4   & \multicolumn{4}{c}{[{\it -0.0031, 0.0432}]} \\
				RME1-STB1   & \multicolumn{4}{c}{[{\bf  0.0000}, 0.0372]} & MCM1-NDD1   & \multicolumn{4}{c}{[{\it -0.0128, 0.0083}]} \\
				SWI4-SWI6   & \multicolumn{4}{c}{[{\bf  0.0000}, 0.0753]} & FKH2-NDD1   & \multicolumn{4}{c}{[{\it -0.0004, 0.0015}]} \\
				HAP4-PDR1   & \multicolumn{4}{c}{[{\bf  0.0000}, 0.0134]} & MSN1-PUT3   & \multicolumn{4}{c}{[{\it -0.0036, 0.0033}]} \\
				ASH1-PHD1   & \multicolumn{4}{c}{[{\bf  0.0000}, 0.0154]} & All others   & \multicolumn{4}{c}{[{\bf  0.0000},{\bf  0.0000}]} \\
				\hline
			\end{tabular}
		}
	\end{table}

We compare our plausible covariates from our SSCI with the literature in Table \ref{table:comparison}. As we can see, most of the identified covariates by our SSCI have been confirmed by the previous studies. All of our identified individual TFs, MBP1, FKH2, NDD1, and SWI6, are experimentally verified \citep{wang_group_2007}. Besides, 12 of our identified cooperative TF pairs have been documented in the literature (MBP1-SWI6, GAT3-PDR1, GAT3-MSN4, MSN4-YAP5, MSN4-PDR1, SWI4-SWI6, HAP4-PDR1, GAL4-RGM1, HIR1-HIR2, MCM1-NDD1, FKH2-NDD1, FKH2-MCM1). The remaining 22 plausible TF pairs have either one or both TFs selected in the literature \citep{wang_group_2007} except the PDR1-RAP1 pair. We use * to indicate that the cooperative TF pairs have either one or both TFs selected in the literature. 
	
For these plausible covariates, one advantage of SSCI is that it reports the possible signs of their effects if one of their boundaries is at zero. For example, [0,xxx] indicates plausibly positive, and [-xxx,0] indicates plausibly negative. We report the estimated signs of TFs as ``plausibly +/-'' in Table \ref{table:comparison}, and further compare the estimated signs with the documented regulatory effects in the literature. Most of the identified TFs and TF pairs (MBP1, FKH2, NDD1, SWI6, MBP1-SWI6, MSN4-YAP5,405MSN4-PDR1, SWI4-SWI6, and HAP4-PDR1) are consistent with the literature \citep{tsai_statistical_2005}.

%\begin{comment}
\begin{table}[H]
\normalsize
\centering
\caption{Real data example: Yeast cell-cycle (G1) data}\label{table:comparison}
\scalebox{0.45}{
\begin{tabular}{cccccccccccccc}
	\toprule
    & \multicolumn{13}{c}{\Large $\text{SSCI}_{0.95}$}  \\
	\cmidrule{2-14}  
\Large $\text{SSCI}_{0.95}$ (SPSP+AdaLasso) & \multicolumn{4}{c}{\Large Confidence intervals} & \multicolumn{5}{c}{\Large Literature evidence} & \multicolumn{4}{c}{\Large Effects documented in the literature} \\
	\midrule	
\Large time        & \multicolumn{4}{c}{ \Large Significantly +++ } & \multicolumn{5}{c}{ \Large } & \multicolumn{4}{c}{ \Large } \\
\Large MBP1        & \multicolumn{4}{c}{ \Large Significantly +++ } & \multicolumn{5}{c}{ \Large \citet{tsai_statistical_2005}, \citet{wang_group_2007}, \citet{yue_two-step_2019}} & \multicolumn{4}{c}{ \Large  + }  \\
\Large SWI6        & \multicolumn{4}{c}{ \Large Plausibly +} & \multicolumn{5}{c}{ \Large \citet{wang_group_2007}, \citet{yue_two-step_2019}} & \multicolumn{4}{c}{ \Large  + } \\
\Large FKH2        & \multicolumn{4}{c}{ \Large Plausibly -} & \multicolumn{5}{c}{ \Large \citet{tsai_statistical_2005}, \citet{wang_group_2007}, \citet{yue_two-step_2019}} & \multicolumn{4}{c}{ \Large  - } \\
\Large NDD1        & \multicolumn{4}{c}{ \Large Plausibly -} & \multicolumn{5}{c}{ \Large \citet{tsai_statistical_2005}, \citet{wang_group_2007}, \citet{yue_two-step_2019}} & \multicolumn{4}{c}{ \Large  - } \\ 
\Large MBP1-SWI6   & \multicolumn{4}{c}{ \Large Plausibly +} & \multicolumn{5}{c}{ \Large \citet{koch_role_1993}, \citet{banerjee_identifying_2003}, \citet{tsai_statistical_2005}} & \multicolumn{4}{c}{ \Large  + } \\ 
\Large RAP1-YAP5   & \multicolumn{4}{c}{ \Large Plausibly +} & \multicolumn{5}{c}{ \Large *} & \multicolumn{4}{c}{ \Large NA} \\ 
\Large GAT3-RAP1   & \multicolumn{4}{c}{ \Large Plausibly +} & \multicolumn{5}{c}{ \Large *} & \multicolumn{4}{c}{ \Large NA} \\ 
\Large PDR1-RAP1   & \multicolumn{4}{c}{ \Large Plausibly +} & \multicolumn{5}{c}{ \Large NA} & \multicolumn{4}{c}{ \Large NA} \\ 

\Large MSN4-RGM1   & \multicolumn{4}{c}{ \Large Plausibly +} & \multicolumn{5}{c}{ \Large *} & \multicolumn{4}{c}{ \Large NA} \\ 
\Large GAT3-RGM1   & \multicolumn{4}{c}{ \Large Plausibly +} & \multicolumn{5}{c}{ \Large *} & \multicolumn{4}{c}{ \Large NA} \\ 
\Large MSN4-YAP5   & \multicolumn{4}{c}{ \Large Plausibly +} & \multicolumn{5}{c}{ \Large \citet{banerjee_identifying_2003}, \citet{tsai_statistical_2005}} & \multicolumn{4}{c}{ \Large +} \\
\Large MSN4-RAP1   & \multicolumn{4}{c}{ \Large Plausibly +} & \multicolumn{5}{c}{ \Large *} & \multicolumn{4}{c}{ \Large NA} \\

\Large GAT3-YAP5   & \multicolumn{4}{c}{ \Large Plausibly +} & \multicolumn{5}{c}{ \Large *} & \multicolumn{4}{c}{ \Large NA} \\
\Large MSN4-PDR1   & \multicolumn{4}{c}{ \Large Plausibly +} & \multicolumn{5}{c}{ \Large \citet{tsai_statistical_2005}} & \multicolumn{4}{c}{ \Large +} \\
\Large PDR1-RGM1   & \multicolumn{4}{c}{ \Large Plausibly +} & \multicolumn{5}{c}{ \Large *} & \multicolumn{4}{c}{ \Large NA} \\
\Large PDR1-YAP5   & \multicolumn{4}{c}{ \Large Plausibly +} & \multicolumn{5}{c}{ \Large *} & \multicolumn{4}{c}{ \Large NA} \\
\Large RME1-STB1   & \multicolumn{4}{c}{ \Large Plausibly +} & \multicolumn{5}{c}{ \Large *} & \multicolumn{4}{c}{ \Large NA} \\
\Large SWI4-SWI6   & \multicolumn{4}{c}{ \Large Plausibly +} & \multicolumn{5}{c}{ \Large \citet{koch_role_1993}, \citet{banerjee_identifying_2003}, \citet{tsai_statistical_2005}} & \multicolumn{4}{c}{ \Large +} \\

\Large HAP4-PDR1   & \multicolumn{4}{c}{ \Large Plausibly +} & \multicolumn{5}{c}{ \Large \citet{tsai_statistical_2005}} & \multicolumn{4}{c}{ \Large +} \\
\Large ASH1-PHD1   & \multicolumn{4}{c}{ \Large Plausibly +} & \multicolumn{5}{c}{ \Large *} & \multicolumn{4}{c}{ \Large NA} \\
\Large GAL4-PDR1   & \multicolumn{4}{c}{ \Large Plausibly +} & \multicolumn{5}{c}{ \Large *} & \multicolumn{4}{c}{ \Large NA} \\
\Large GAL4-RGM1   & \multicolumn{4}{c}{ \Large Plausibly +} & \multicolumn{5}{c}{ \Large \citet{banerjee_identifying_2003}} & \multicolumn{4}{c}{ \Large NA} \\
\Large HAL9-MAC1   & \multicolumn{4}{c}{ \Large Plausibly +} & \multicolumn{5}{c}{ \Large *} & \multicolumn{4}{c}{ \Large NA} \\
\Large MAC1-NRG1   & \multicolumn{4}{c}{ \Large Plausibly +} & \multicolumn{5}{c}{ \Large *} & \multicolumn{4}{c}{ \Large NA} \\

\Large HAP4-MSN4   & \multicolumn{4}{c}{ \Large Plausibly +} & \multicolumn{5}{c}{ \Large *} & \multicolumn{4}{c}{ \Large NA} \\
\Large HAP4-MET31  & \multicolumn{4}{c}{ \Large Plausibly +} & \multicolumn{5}{c}{ \Large *} & \multicolumn{4}{c}{ \Large NA} \\
\Large GLN3-MET31  & \multicolumn{4}{c}{ \Large Plausibly +} & \multicolumn{5}{c}{ \Large *} & \multicolumn{4}{c}{ \Large NA} \\
\Large MET31-ZMS1  & \multicolumn{4}{c}{ \Large Plausibly +} & \multicolumn{5}{c}{ \Large *} & \multicolumn{4}{c}{ \Large NA} \\
\Large NRG1-RLM1   & \multicolumn{4}{c}{ \Large Plausibly +} & \multicolumn{5}{c}{ \Large *} & \multicolumn{4}{c}{ \Large NA} \\
\Large HIR1-HIR2   & \multicolumn{4}{c}{ \Large Plausibly +} & \multicolumn{5}{c}{ \Large \citet{loy_ndd1_1999}, \citet{banerjee_identifying_2003}} & \multicolumn{4}{c}{ \Large NA} \\
\Large STP1-STP2   & \multicolumn{4}{c}{ \Large Plausibly +} & \multicolumn{5}{c}{ \Large *} & \multicolumn{4}{c}{ \Large NA} \\
\Large FKH2-MCM1   & \multicolumn{4}{c}{ \Large Plausibly -} & \multicolumn{5}{c}{ \Large \citet{kumar_forkhead_2000}, \citet{banerjee_identifying_2003}} & \multicolumn{4}{c}{ \Large NA} \\
\Large GCR2-SRD1   & \multicolumn{4}{c}{ \Large Plausibly -} & \multicolumn{5}{c}{ \Large *} & \multicolumn{4}{c}{ \Large NA} \\
\Large GAT3-PDR1   & \multicolumn{4}{c}{ \Large Plausibly} & \multicolumn{5}{c}{ \Large \citet{banerjee_identifying_2003}, \citet{tsai_statistical_2005}} & \multicolumn{4}{c}{ \Large +} \\ 
\Large GAT3-MSN4   & \multicolumn{4}{c}{ \Large Plausibly} & \multicolumn{5}{c}{ \Large \citet{banerjee_identifying_2003}} & \multicolumn{4}{c}{ \Large NA} \\
\Large MCM1-NDD1   & \multicolumn{4}{c}{ \Large Plausibly} & \multicolumn{5}{c}{ \Large \citet{kumar_forkhead_2000} \citet{banerjee_identifying_2003}, \citet{tsai_statistical_2005}} & \multicolumn{4}{c}{ \Large -} \\
\Large FKH2-NDD1   & \multicolumn{4}{c}{ \Large Plausibly} & \multicolumn{5}{c}{ \Large \citet{kumar_forkhead_2000}, \citet{banerjee_identifying_2003}, \citet{tsai_statistical_2005}} & \multicolumn{4}{c}{ \Large NA} \\ 
\Large MSN1-PUT3   & \multicolumn{4}{c}{ \Large Plausibly} & \multicolumn{5}{c}{ \Large *} & \multicolumn{4}{c}{ \Large NA} \\

	\midrule
	& \multicolumn{13}{c}{\Large \text{MCB}_{0.95}}  \\
	\cmidrule{2-14}  
	& \multicolumn{6}{c}{\Large Lower confidence bound model} & \multicolumn{7}{c}{\Large Upper confidence bound model}  \\
	\cmidrule(r){2-7}  \cmidrule(l){8-14}  
	\Large SSCI (SPSP+AdaLasso) & \multicolumn{6}{c}{\Large $\{\text{time, MBP1}\}$} & \multicolumn{7}{c}{\Large $\{\text{time, MBP1, and all plausible covariates above}\}$}  \\
	\bottomrule
	\multicolumn{14}{@{}p{35cm}@{}}{\large Note: We provide the literature evidence for our selected cooperative TF pairs. We use * to represent the cooperative TF pairs with either one or both TFs of the pair selected in the literature. We display ``Significantly positive/negative'' as ``Significantly +++/- - -'', and ``Plausibly positive/negative'' as ``Plausibly +/-''. The last column shows the sign of the coefficients documented in the corresponding literature.} \\
	
 \multicolumn{14}{r}{We use * to represent the cooperative TF pairs that have either one or both TFs of the pair selected in the literature. We display ``Significantly positive (negative)'' as ``Significantly +++($---$)'',} \\ 
 \multicolumn{14}{l}{and ``Plausibly positive/negative'' as ``Plausibly +/-''.} \\ 
\end{tabular}
}
\end{table}

\newpage
\section{Technical Proofs}\label{Proofs}	
Assume the regression model (1). Without loss of generality, we assume the $\mathbf{X}$ are standardized with zero mean, unit variance. 
	
	\begin{lemma}\label{lemma_allcorrect}
		Suppose the selection method selects the true model with the probability $1-2e^{-h(n)}$, that is $\mathbf{P}(\tilde{\mathcal{S}}=\mathcal{S}_{0})=1-2e^{-h(n)}$, then $\mathbf{P}\left(\underset{b \in B}{\cap}\left(\hat{\mathcal{S}}^{(b)}=\tilde{\mathcal{S}}\right), \tilde{\mathcal{S}}=\mathcal{S}_{0}\right) \xrightarrow{n\rightarrow\infty} 1$ holds under the condition $B=o(e^{h(n)})$.
	\end{lemma}

	\begin{proof} 
		\textit{Proof.} With a selection method that can select the true model with the probability $1-2e^{-h(n)}$, we have \begin{align*}
			\; & \; \mathbf{P}\left(\underset{b \in B}{\cap}\left(\hat{\mathcal{S}}^{(b)}=\tilde{\mathcal{S}}\right), \tilde{\mathcal{S}}=\mathcal{S}_{0}\right) \\
			= \; & \; \mathbf{P}\left(\underset{b \in B}{\cap}\left(\hat{\mathcal{S}}^{(b)}=\tilde{\mathcal{S}}\right) \middle| \tilde{\mathcal{S}}=\mathcal{S}_{0}\right)\mathbf{P}\left(\tilde{\mathcal{S}}=\mathcal{S}_{0}\right) \\
			\ge \; & \; \left(1-2e^{-h(n)}\right)^B \left(1-2e^{-h(n)}\right) \\
			= \; & \; \left(1-2e^{-h(n)}\right)^{B+1} = \text{exp}\left((B+1)\log\left(1-2e^{-h(n)}\right)\right) .
		\end{align*}
		
		By Taylor expansion, \begin{align*}
			\underset{n\rightarrow\infty}{\text{lim}}\Bigg((B+1)\log\left(1-2e^{-h(n)}\right)\Bigg) \; & = \; \underset{n\rightarrow\infty}{\text{lim}}\Bigg(\left(B+1\right)\Big[-2e^{-h(n)} + o\left(2e^{-h(n)}\right) \Big]\Bigg)\\
			\; & = \; \underset{n\rightarrow\infty}{\text{lim}}\Bigg(-2\left(B+1\right)e^{-h(n)} + o\left(2(B+1)e^{-h(n)}\right) \Big]\Bigg)\\
		\end{align*} 
		
		Under condition $B=o(e^{h(n)})$, \begin{align*}
			\underset{n\rightarrow\infty}{\text{lim}}\Bigg(\left({B}+1\right)\log\left(1-2e^{-h(n)}\right)\Bigg) \; & \; = \underset{n\rightarrow\infty}{\text{lim}}\Bigg(-2({B}+1)e^{-h(n)}\Bigg) \\
			\; & \; =\underset{n\rightarrow\infty}{\text{lim}}\left(-\frac{\big[2({B}+1)\big]}{e^{h(n)}}\right) \\
			\; & \; = 0. \\
			\text{Therefore,} \;\; \mathbf{P}(\underset{b \in B}{\cap}(\hat{\mathcal{S}}^{(b)}=\tilde{\mathcal{S}}), \tilde{\mathcal{S}}=\mathcal{S}_{0}) \; & \ge \; \;\; (1-2e^{-h(n)})^{B+1} \\
			\; &  = \; \;\; e^{({B}+1)\log(1-2e^{-h(n)})} \\
			\; &  \overset{n\rightarrow\infty}{\rightarrow}  1.
		\end{align*}
	\end{proof}
	
	\begin{lemma}\label{lemma_dist}
		Suppose the selection method selects the true model with the probability $1-2e^{-h(n)}$, and $h(n)/\log(n+1) \rightarrow \infty$, the $\mathbf{X}_{\mathcal{S}_0}^T\mathbf{X}_{\mathcal{S}_0}/n \rightarrow V_{\mathcal{S}_0}$ and $V_{\mathcal{S}_0}$ is positive definite, then the conditional distribution of $\sqrt{n} \left( \hat{\boldsymbol{\beta}}^{(b)} - \tilde{\boldsymbol{\beta}} \right)$ converges to normal with mean $\mathbf{0}$ and variance-covariance matrix $
		\begin{bmatrix} 
			\sigma^2 V_{\mathcal{S}_{0}}^{-1} & \mathbf{0}_{s_0\times (p-s_0)} \\
			\mathbf{0}_{(p-s_0)\times s_0} & \mathbf{0}_{(p-s_0)\times (p-s_0)} \\
		\end{bmatrix}
		\quad$ as $n \rightarrow \infty$.
	\end{lemma}
	
	\begin{proof} 
		\textit{Proof.} Suppose $\hat{\boldsymbol{\beta}}^{\text{LS}}_{\mathcal{S}_{0}} \in \mathbb{R}^{s_0}$ and $\hat{\boldsymbol{\beta}}^{\text{LS}(b)}_{\mathcal{S}_{0}} \in \mathbb{R}^{s_0}$ are the least square estimators when the true model $\mathcal{S}_{0}$ is known. Then we denote  $\hat{\boldsymbol{\beta}}_{\mathcal{S}_{0}}= \left(\hat{\boldsymbol{\beta}}^{\text{LS}}_{\mathcal{S}_{0}}, \mathbf{0}\right) \in \mathbb{R}^p$ and $\hat{\boldsymbol{\beta}}^{(b)}_{\mathcal{S}_{0}}= \left(\hat{\boldsymbol{\beta}}^{\text{LS}(b)}_{\mathcal{S}_{0}}, \mathbf{0}\right) \in \mathbb{R}^p$. We decompose $\sqrt{n} \left( \hat{\boldsymbol{\beta}}^{(b)} - \tilde{\boldsymbol{\beta}} \right)$ as follows: $$\sqrt{n} \left( \hat{\boldsymbol{\beta}}^{(b)} - \tilde{\boldsymbol{\beta}} \right) =\sqrt{n} \left( \hat{\boldsymbol{\beta}}^{(b)} - \hat{\boldsymbol{\beta}}^{(b)}_{\mathcal{S}_{0}} \right) + \sqrt{n} \left(  \hat{\boldsymbol{\beta}}^{(b)}_{\mathcal{S}_{0}} -  \hat{\boldsymbol{\beta}}_{\mathcal{S}_{0}} \right) + \sqrt{n} \left( \hat{\boldsymbol{\beta}}_{\mathcal{S}_{0}} - \tilde{\boldsymbol{\beta}} \right) .$$ 
		By Lemma \ref{lemma_allcorrect}, the first and third items tend to zero in probability, that is $\sqrt{n} \left( \hat{\boldsymbol{\beta}}^{(b)} - \tilde{\boldsymbol{\beta}} \right) = \sqrt{n} \left(  \hat{\boldsymbol{\beta}}^{(b)}_{\mathcal{S}_{0}} -  \hat{\boldsymbol{\beta}}_{\mathcal{S}_{0}} \right) + o_p(1).$
		
		Therefore, by Slutsky's theorem, and the residual bootstrap validity in \citet{freedman_bootstrapping_1981}, the conditional distribution of $\sqrt{n} \left( \hat{\boldsymbol{\beta}}^{(b)} - \tilde{\boldsymbol{\beta}} \right)$ converges to the distribution of $\sqrt{n} \left(\hat{\boldsymbol{\beta}}_{\mathcal{S}_{0}} -  \boldsymbol{\beta}^0 \right)$ that is a normal distribution with mean $\mathbf{0}$ and covariance matrix $\begin{bmatrix} 
			\sigma^2 V_{\mathcal{S}_{0}}^{-1} & \mathbf{0}_{s_0\times (p-s_0)} \\
			\mathbf{0}_{(p-s_0)\times s_0} & \mathbf{0}_{(p-s_0)\times (p-s_0)} \\
		\end{bmatrix}$. 
	\end{proof}
	
	\begin{theorem} \label{theorem_SCI} Suppose the selection method selects the true model with the probability $1-2e^{-h(n)}$, that is $\mathbf{P}(\tilde{\mathcal{S}}=\mathcal{S}_{0})=1-2e^{-h(n)}$. Then, with $B=o(e^{h(n)})$, for any confidence level of $\alpha\in (0,1)$, the $\textup{SSCI}_{1-\alpha}$ constructed in Algorithm 1 has the asymptotic coverage probability  $\mathbf{P}(\boldsymbol{\beta}^0 \in \textup{SSCI}_{1-\alpha}) \xrightarrow{n\rightarrow\infty} 1-\alpha.$
    \end{theorem}

	\begin{proof} \textit{Proof.} The coverage probability is equivalent to: \begin{align*} \label{eqn:2.1}
			\; & \; \mathbf{P}\left(\boldsymbol{\beta}^{0}\in \text{SSCI}_{(1-\alpha)}\right) \\
			= \; & \; \mathbf{P}\left(\boldsymbol{\beta}^{0}\in \text{SSCI}_{(1-\alpha)}, \underset{b \in B}{\cap}\left(\hat{\mathcal{S}}^{(b)}=\tilde{\mathcal{S}}\right), \tilde{\mathcal{S}}=\mathcal{S}_{0}\right) \\
			= \; & \; \mathbf{P}\left(\boldsymbol{\beta}^{0}\in \text{SSCI}_{(1-\alpha)} \middle| \underset{b \in B}{\cap}\left(\hat{\mathcal{S}}^{(b)}=\tilde{\mathcal{S}}\right), \tilde{\mathcal{S}}=\mathcal{S}_{0}\right)\mathbf{P}\left(\underset{b \in B}{\cap}\left(\hat{\mathcal{S}}^{(b)}=\tilde{\mathcal{S}}\right),\tilde{\mathcal{S}}=\mathcal{S}_{0}\right)  \tag{2.1}  \end{align*} 
		%\end{equation}
		
		By Lemma \ref{lemma_dist}, the conditional distribution of $\sqrt{n}(\hat{\boldsymbol{\beta}}^{(b)}-\tilde{\boldsymbol{\beta}})$ converges weakly to normal with mean $\mathbf{0}$ and variance-covariance $
		\begin{bmatrix} 
			\sigma^2 V_{\mathcal{S}_{0}}^{-1} & \mathbf{0}_{s_0\times (p-s_0)} \\
			\mathbf{0}_{(p-s_0)\times s_0} & \mathbf{0}_{(p-s_0)\times (p-s_0)} \\
		\end{bmatrix} ,
		\quad
		$ which is symmetric. In general, for symmetric distribution, \begin{equation}\label{eqn:boots}
			\mathbf{P}\left(\boldsymbol{a} \le \tilde{\boldsymbol{\beta}}-\boldsymbol{\beta}^{0} \le \boldsymbol{b}\right) = \mathbf{P}\left(-\boldsymbol{a} \ge \tilde{\boldsymbol{\beta}}-\boldsymbol{\beta}^{0} \ge -\boldsymbol{b}\right) = \mathbf{P}\left(-\boldsymbol{b} \le \tilde{\boldsymbol{\beta}}-\boldsymbol{\beta}^{0} \le -\boldsymbol{a}\right).
			\tag{2.2} 
		\end{equation} % In addition, since $\hat{\boldsymbol{\beta}}^{(b)}-\tilde{\boldsymbol{\beta}}$ and $\tilde{\boldsymbol{\beta}}-\boldsymbol{\beta}^{0}$ have the same asymptotic distribution, 
		If we construct the simultaneous confidence intervals $\left(\underline{\boldsymbol{\beta}}, \overline{\boldsymbol{\beta}}\right)$ for $\hat{\boldsymbol{\beta}}^{(b)}$ such that $\mathbf{P}\left(\underline{\boldsymbol{\beta}}\le \hat{\boldsymbol{\beta}}^{(b)} \le \overline{\boldsymbol{\beta}}\right) = 1- \alpha$. Then, \begin{align*}
			\; & \; \mathbf{P}\left(\underline{\boldsymbol{\beta}} \le \hat{\boldsymbol{\beta}}^{(b)}\le \overline{\boldsymbol{\beta}}\right) \\
			= \; & \; \mathbf{P}\left(\underline{\boldsymbol{\beta}} - \tilde{\boldsymbol{\beta}}  \le \hat{\boldsymbol{\beta}}^{(b)}- \tilde{\boldsymbol{\beta}} \le \overline{\boldsymbol{\beta}} - \tilde{\boldsymbol{\beta}}\right) \\
			\; & \; \left( \tilde{\boldsymbol{\beta}}-\boldsymbol{\beta}^{0} \text{ and } \hat{\boldsymbol{\beta}}^{(b)}-\tilde{\boldsymbol{\beta}} \text{ have the same symmetric distribution} \right) \\
			= \; & \; \mathbf{P}\left(\underline{\boldsymbol{\beta}} - \tilde{\boldsymbol{\beta}}  \le \tilde{\boldsymbol{\beta}}- \boldsymbol{\beta}^{0} \le  \overline{\boldsymbol{\beta}} - \tilde{\boldsymbol{\beta}}\right) \; \;\;\; \qquad \left(\text{apply \eqref{eqn:boots}  due to the symmetry of } \tilde{\boldsymbol{\beta}} - \boldsymbol{\beta}^{0}\right)\\
			= \; & \; \mathbf{P}\left(\tilde{\boldsymbol{\beta}}-\overline{\boldsymbol{\beta}} \le \tilde{\boldsymbol{\beta}} - \boldsymbol{\beta}^{0} \le  \tilde{\boldsymbol{\beta}} -\underline{\boldsymbol{\beta}}\right) \; \;\;\; \qquad \left(\text{multiply inside term by -1}\right)  \\
			= \; & \; \mathbf{P}\left(\underline{\boldsymbol{\beta}} - \tilde{\boldsymbol{\beta}} \le \boldsymbol{\beta}^{0} - \tilde{\boldsymbol{\beta}} \le \overline{\boldsymbol{\beta}} - \tilde{\boldsymbol{\beta}}\right) \\
			= \; & \; \mathbf{P}\left(\underline{\boldsymbol{\beta}} \le \boldsymbol{\beta}^{0} \le \overline{\boldsymbol{\beta}}\right)\\
			= \; & \; 1-\alpha. \; \;\;\; \qquad \left(\text{construction of } \left(\underline{\boldsymbol{\beta}}, \overline{\boldsymbol{\beta}}\right)\right) \\
		\end{align*} 
		
		By the above property, the first term of \eqref{eqn:2.1} is $\mathbf{P}\left(\boldsymbol{\beta}^{0}\in \text{SSCI}_{(1-\alpha)} \middle| \underset{b \in B}{\cap}\left(\hat{\mathcal{S}}^{(b)}=\tilde{\mathcal{S}}=\mathcal{S}_{0}\right)\right) = 1-\alpha.$
		
		Use Lemma \ref{lemma_allcorrect} for the second term of \eqref{eqn:2.1}, the coverage probability of our SSCI is	\begin{align*}
			& \; \mathbf{P}\left(\boldsymbol{\beta}^{0}\in \text{SSCI}_{(1-\alpha)}\right) \\
			= \; \; & \; \mathbf{P}\left(\boldsymbol{\beta}^{0}\in \text{SSCI}_{(1-\alpha)} \middle| \underset{b \in B}{\cap}\left(\hat{\mathcal{S}}^{(b)}=\tilde{\mathcal{S}}\right), \tilde{\mathcal{S}}=\mathcal{S}_{0}\right)\mathbf{P}\left(\underset{b \in B}{\cap}\left(\hat{\mathcal{S}}^{(b)}=\tilde{\mathcal{S}}\right),\tilde{\mathcal{S}}=\mathcal{S}_{0}\right) \\
			\overset{n\rightarrow\infty}{\longrightarrow} \; & \; 1-\alpha.
		\end{align*} 
	\end{proof}

\begin{corollary}[Lasso]\label{corollary_Lasso}
Let $\tilde{\mathcal{S}}^{\textup{Lasso}}(\lambda_n)$ be the model selected by the Lasso with the tuning parameter $\lambda_n$ and $\textup{SSCI}^{\textup{Lasso}}_{1-\alpha} (\lambda_n)$ be constructed by Algorithm 1 using the Lasso with the tuning parameter $\lambda_n$ and the least square refitted estimate. Under the strong irrepresentable condition \citep{zhao_model_2006}, $\mathbf{P}(\tilde{\mathcal{S}}^{\textup{Lasso}}(\lambda_n)=\mathcal{S}_{0}) \ge 1-2e^{-n^c}$ if $\lambda_n/n \rightarrow 0$ and $\lambda_n/n^{(1+c)/2} \rightarrow \infty$ with $0\le c < 1$, and $B=o(e^{n^c})$ , we have
$\mathbf{P}(\boldsymbol{\beta}^0 \in \textup{SSCI}^{\textup{Lasso}}_{1-\alpha}(\lambda_n)) \xrightarrow{n\rightarrow\infty} 1-\alpha.$
\end{corollary}

	\begin{proof} \textit{Proof.} Under the strong irrepresentable condition in the study of \citet{zhao_model_2006} and $B=o(e^{n^c})$ with $0\le c < 1$. With $\lambda_n/n \rightarrow 0$ and $\lambda_n/n^{({(1+c)}/{2})} \rightarrow \infty$, $\mathbf{P}(\tilde{\mathcal{S}}^{Lasso}(\lambda_n)=\mathcal{S}_{0}) \ge (1-2e^{-n^c})$. 
		
		Similar to the Proof in Lemma \ref{lemma_allcorrect}:  \begin{align*}
			\mathbf{P}\left(\underset{b \in B}{\cap}\left(\hat{\mathcal{S}}^{(b)}_{\text{Lasso}}=\tilde{\mathcal{S}}^{\text{Lasso}}\right), \tilde{\mathcal{S}}^{\text{Lasso}}=\mathcal{S}_{0}\right) & \ge  \left(1-2e^{-n^c}\right)^{B+1} \\
			& = e^{(B+1)\log(1-2e^{-n^c})} \\
			& \overset{n\rightarrow\infty}{\longrightarrow}  1.
		\end{align*}
		
		Therefore, \[\mathbf{P}\left(\boldsymbol{\beta}\in \text{SSCI}^{Lasso}_{(1-\alpha)}\left(\lambda_n\right)\right) \xrightarrow{n\rightarrow\infty} 1-\alpha.\]
	\end{proof}

\begin{corollary}[Adaptive Lasso]\label{corollary_AdaLasso}
Let $\tilde{\mathcal{S}}^{\textup{Ada}}(\lambda_n)$ be the model selected by the adaptive Lasso with the tuning parameter $\lambda_n$ and $\textup{SSCI}^{\textup{Ada}}_{1-\alpha} (\lambda_n)$ be constructed by Algorithm 1 using the adaptive Lasso with the tuning parameter $\lambda_n$ and the least square refitted estimate. Under the restricted eigenvalue condition \citep{geer_adaptive_2011},  $\mathbf{P}(\tilde{\mathcal{S}}^{\textup{Ada}}(\lambda_n)=\mathcal{S}_{0})\ge 1-2e^{-\gamma n}$ if  $\lambda_n=4\sigma\sqrt{2\gamma+ 2\log p/n }$ and  $\gamma\rightarrow 0$. Further if $B=o(e^{\gamma n})$, we have
$\mathbf{P}(\boldsymbol{\beta}^0 \in \textup{SSCI}^{\textup{Ada}}_{1-\alpha}(\lambda_n)) \xrightarrow{n\rightarrow\infty} 1-\alpha.$
\end{corollary}

	\begin{proof} \textit{Proof.} Under the restricted eigenvalue condition in \citet{geer_adaptive_2011},  $\mathbf{P}(\tilde{\mathcal{S}}^{\text{Adap}}(\lambda_n)=\mathcal{S}_{0})\ge (1-2e^{-\gamma n})$ if  $\lambda_n=4\sigma\sqrt{2\gamma+ {2\log p}/{n}}$ and  $\gamma\rightarrow 0$. Further if $B=o(e^{\gamma n})$, 
		
		\begin{align*}
			\mathbf{P}\left(\underset{b \in B}{\cap}\left(\hat{\mathcal{S}}^{(b)}_{\text{Adap}}=\tilde{\mathcal{S}}^{\text{Adap}}\right), \tilde{\mathcal{S}}^{\text{Adap}}=\mathcal{S}_{0}\right) \; & \ge \;\;\;  \left(1-2e^{-\gamma n}\right)^{B+1} \\
			& = \;\;\; e^{(B+1)\log(1-2e^{-\gamma n})} \\
			& \overset{n\rightarrow\infty}{\longrightarrow}  1.
		\end{align*}
		
		Therefore,
		\[\mathbf{P}\left(\boldsymbol{\beta}\in \text{SSCI}^{\text{Adap}}_{(1-\alpha)}(\lambda_n)\right) \xrightarrow{n\rightarrow\infty} 1-\alpha.\]
	\end{proof}

\begin{corollary}[SPSP]\label{corollary_SPSP}
Let $\tilde{\mathcal{S}}^{\textup{SPSP}}$ be the model selected by SPSP and $\textup{SSCI}^{\textup{SPSP}}_{1-\alpha}$ be constructed by Algorithm 1 using the SPSP and the least square refitted estimate. Under the compatibility condition in \citet{buhlmann_statistics_2011}, and the weak identifiability condition in \citet{liu_selection_2018}, the SPSP can select the true model $\mathcal{S}_{0}$ over $\lambda\in [4\sigma\sqrt{2\gamma + {2\log p}/{n}},+\infty]$ and $\gamma \rightarrow 0$ with probability at least $1-2e^{-\gamma n}$, that is $\mathbf{P}(\tilde{\mathcal{S}}^{\textup{SPSP}}=\mathcal{S}_{0})\ge (1-2e^{-\gamma n})$. Further if $B=o(e^{\gamma n})$, we have
$\mathbf{P} (\boldsymbol{\beta}^0 \in \textup{SSCI}^{\textup{SPSP}}_{1-\alpha} ) \xrightarrow{n\rightarrow\infty} 1-\alpha.$
\end{corollary}

	\begin{proof} \textit{Proof.} Under the compatibility condition in \citet{van_de_geer_deterministic_2007} and \citet{geer_adaptive_2011}, and the weak identifiability condition in \citet{liu_selection_2018}, the SPSP can select the true model $\mathcal{S}_{0}$ over $\lambda\in \Big[4\sigma\sqrt{2\gamma + {2\log p}/{n}},+\infty\Big]$ with probability at least $(1-2e^{-\gamma n})$, that is $\mathbf{P}(\tilde{\mathcal{S}}^{\text{SPSP}}=\mathcal{S}_{0})\ge (1-2e^{-\gamma n})$. Further if $B=o(e^{\gamma n})$,
		
		\begin{align*}
			\mathbf{P}\left(\underset{b \in B}{\cap}\left(\hat{\mathcal{S}}^{(b)}_{\text{SPSP}}=\tilde{\mathcal{S}}^{\text{SPSP}}\right), \tilde{\mathcal{S}}^{\text{SPSP}}=\mathcal{S}_{0}\right) & \ge \;\;\; \left(1-2e^{-\gamma n}\right)^{B+1} \\
			& = \;\;\; e^{(B+1)\log(1-2e^{-\gamma n})} \\
			& \overset{n\rightarrow\infty}{\longrightarrow}  1.
		\end{align*}
		
		Therefore, \[\mathbf{P} \left(\boldsymbol{\beta}\in \text{SSCI}^{\text{SPSP}}_{(1-\alpha)}\right) \xrightarrow{n\rightarrow\infty} 1-\alpha.\]
	\end{proof}

\newpage

\section{Additional Simulation Studies}\label{Proofs}	

As per request from the reviewers, we add two more simulation settings with higher p/n ratio and a large number of non-zero regression coefficients in addition to the existing simulation setups. 
From the results of three additional setups, our SSCIs can still perform well under these scenarios. 
The coverage rates are still close to the nominal 95\%. 
Moreover, the average interval width of true signals $\bar{w}_{\mathcal{S}_{0}}$ and non-signals $\bar{w}_{\mathcal{S}_{0}^c}$ are still smaller than the best alternative method SCI (Debiased Lasso) proposed by \citet{zhang_simultaneous_2017}. 
In Example 10, the average width of intervals from SCI (Debiased Lasso) is very wide, as wide as three times of the width of our SSCI based on the SPSP selection methods. 
Therefore, we emphasize that our SSCI is a practically very useful inference tool that can effectively provide more insights about the true model or what covariates could be confidently removed as irrelevant ones.
We will include these new simulation studies in our supplementary document for our paper. The simulation settings and results are presented as follows: 

{\bf Example 8:} Let $n=50$, $p=300$, the $p/n=6$ and $\boldsymbol{\beta}^{0}=(3,2, 2,0,...,0)$. The covariates are independent. 

{\bf Example 9:} Let $n=50$, $p=150$, the $p/n=3$ and $\boldsymbol{\beta}^{0}=(3,2,1.5,0,...,0)$. The pairwise covariate correlation is $\text{cor}(\mathbf{X}_j, \mathbf{X}_{j'})=0.5^{|j - {j'}|}$. 

{\bf Example 10:} Let $n = 100$, $p = 150$,  $\boldsymbol{\beta}^{0}=(2,2,2,2,2,2,2,2,2,2,2,2,2,2,2,...,0)$. In total, there are $s_0=15$ non-zero regression coefficients. The covariates are independent.

%\begin{singlespace}
\begin{table}[H]
	\renewcommand{\arraystretch}{0.8} % this reduces the vertical spacing between rows
	\centering
	\caption{Simulation results of Examples 8, 9, \& 10}\label{table:simE8}
%	\vspace{-5mm}
	\scalebox{0.65}{
		
%		{\color{blue}
			\begin{tabular}{@{}cl@{\extracolsep{5pt}} cccrcc}
				\multicolumn{8}{@{}p{20cm}@{}}{\footnotesize } \\
				\toprule
%				\multicolumn{8}{c}{Study 1: High-dimensional settings} \\
				& & \multicolumn{3}{c}{Simultaneous Confidence Intervals} & \multicolumn{3}{c}{Model Confidence Bounds} \\
				\cmidrule{3-5}  \cmidrule{6-8}  
				Setting & Method &  $\text{P}^{\text{SCI}}_{\text{coverage}}(\%)$  & $\bar{w}_{\mathcal{S}_0}$ & $\bar{w}_{\mathcal{S}_0^c}$ & \multicolumn{1}{r}{$\text{P}^{\text{MCB}}_{\text{coverage}}(\%)$} & \multicolumn{2}{c}{$\bar{w}$} \\ 
				\midrule 
				& SSCI (SPSP+SCAD) & 91.1 & 0.692 (0.004) & {\textbf{0.001}} (0.000) & 100.0 & \multicolumn{2}{c}{0.60}   \\ 
				& SSCI (SPSP+MCP) & 91.1 & 0.666 (0.003) & {\textbf{0.000}} (0.000) & 100.0 & \multicolumn{2}{c}{0.05} \\ 
				\hfil Example 8 & SSCI (SPSP+AdaLasso) & 93.2 & 0.715 (0.009) & \textbf{0.001} (0.001) & 99.9 & \multicolumn{2}{c}{0.32}   \\ 
				\hfil $n$=50, $p$=300  & SSCI (SPSP+Lasso) & 93.7 & 0.865 (0.016) & \textbf{0.004} (0.002) & 99.8 & \multicolumn{2}{c}{1.37} \\ 
				\hfil $p/n$=6  & SSCI (SCAD+CV) & 96.0 & 0.891 (0.006) & 0.476 (0.005) & 100.0 & \multicolumn{2}{c}{277.01}   \\ 
				\hfil $s$=3, $\rho$=0 & SSCI (MCP+CV) & 97.2 & 0.907 (0.005) & 0.518 (0.004) & 100.0 & \multicolumn{2}{c}{268.78}   \\ 
				\hfil $MC$=1000, $B$=1000  & SSCI (AdaLasso+CV) & 93.1 & 0.844 (0.006) & 0.213 (0.003) & 99.7 & \multicolumn{2}{c}{148.01}   \\ 
%				& SSCI (Lasso+CV) & 23.9 & 0.838 (0.107) & 0.271 (0.014) & 52.3 & \multicolumn{2}{c}{196.45}   \\ 
				& SCI (hdi) & 0.0 & 0.611 (0.003) & 0.491 (0.003)  & --- &  \multicolumn{2}{c}{---}  \\ 
				& SCI (hdm)$^\ast$ & ---  & ---  & ---   & --- &  \multicolumn{2}{c}{---}  \\ 
				& SCI (Debiased Lasso) & 97.3 & 1.403 (0.006) & 1.403 (0.006)  & --- &  \multicolumn{2}{c}{---}  \\ 
%				& Oracle (Bootstrap) & 89.8 & 0.651 (0.002) & 0.000 (0.000) & --- & \multicolumn{2}{c}{---}  \\ 	
				& Oracle & 94.2 & 0.708 (0.002) & 0.000 (0.000) & --- & \multicolumn{2}{c}{---}  \\ 	
   
                \cmidrule{1-8}
    
%    Example 2-4 n=50, p=300, p/n ratio is 6

                & SSCI (SPSP+SCAD) & 94.9 & 0.919 (0.011) & \textbf{0.002} (0.000) & 99.8 & \multicolumn{2}{c}{0.74}   \\ 
				& SSCI (SPSP+MCP) & 94.9 & 0.862 (0.009) & \textbf{0.000} (0.000) & 100.0 & \multicolumn{2}{c}{0.26} \\ 
				\hfil Example 9 & SSCI (SPSP+AdaLasso) & {{86.0}} & 1.761 (0.025) & {\textbf{0.059}} (0.008) & 93.4 & \multicolumn{2}{c}{7.46}   \\ 
				\hfil $n$=50, $p$=150  & SSCI (SPSP+Lasso) & {{94.5}} & 1.101 (0.014) & {\textbf{0.001}} (0.000) & 99.1 & \multicolumn{2}{c}{0.79} \\ 
				\hfil $p/n$=3 & SSCI (SCAD+CV) & 98.4 & 1.118 (0.009) & 0.574 (0.005) & 100.0 & \multicolumn{2}{c}{138.47}   \\ 
				\hfil $s$=3, $\rho$=0.5 & SSCI (MCP+CV) & 97.2 & 1.109 (0.009) & 0.608 (0.005) & 100.0 & \multicolumn{2}{c}{136.03}   \\ 
				\hfil $MC$=1000, $B$=1000 & SSCI (AdaLasso+CV) & 95.0 & 0.929 (0.005) & 0.166 (0.003) & 100.0 & \multicolumn{2}{c}{51.91}   \\ 
				& SCI (hdi) & 0.1 & 0.711 (0.003) & 0.561 (0.003)  & --- &  \multicolumn{2}{c}{---}  \\ 
				& SCI (hdm) & 37.2 & 1.407 (0.011) & 1.142 (0.005)  & --- &  \multicolumn{2}{c}{---}  \\ 
				& SCI (Debiased Lasso) & 95.8 & 1.257 (0.005) & 1.257 (0.005)  & --- &  \multicolumn{2}{c}{---}  \\ 
				& Oracle & 96.0 & 0.848 (0.004) & 0.000 (0.000) & --- & \multicolumn{2}{c}{---}  \\ 	
                
                \cmidrule{1-8}

%    Example 2-3 n=50, p=150, p/n ratio is 3

				& SSCI (SPSP+SCAD) & 92.2 & 0.674 (0.003) & {\textbf{0.023}} (0.000) & 99.5 & \multicolumn{2}{c}{8.40}   \\ 
				& SSCI (SPSP+MCP) & 92.7 & 0.657 (0.003) & {\textbf{0.039}} (0.002) & 100.0 & \multicolumn{2}{c}{12.90} \\ 
				\hfil Example 10 & SSCI (SPSP+AdaLasso) & 88.5 & 0.574 (0.002) & \textbf{0.001} (0.000) & 100.0 & \multicolumn{2}{c}{0.51}   \\ 
			  \hfil $n$=100, $p$=150  & SSCI (SPSP+Lasso) & 95.0 & 0.687 (0.003) & {\textbf{0.038}} (0.000) & 100.0 & \multicolumn{2}{c}{16.41} \\ 
				\hfil $s$=15, $\rho$=0 & SSCI (SCAD+CV) & 94.5 & 0.686 (0.003) & 0.535 (0.003) & 100.0 & \multicolumn{2}{c}{131.60}   \\ 
			  \hfil $MC$=1000, $B$=2000  & SSCI (MCP+CV) & 95.0 & 0.685 (0.003) & 0.555 (0.003) & 100.0 & \multicolumn{2}{c}{130.80}   \\ 
				& SSCI (AdaLasso+CV) & 92.6 & 0.644 (0.003) & 0.189 (0.003) & 100.0 & \multicolumn{2}{c}{73.83}   \\ 
				& SCI (hdi) & 0.0 & 0.440 (0.002) & 0.308 (0.001)  & --- &  \multicolumn{2}{c}{---}  \\ 
				& SCI (hdm) & 17.3 & 2.124 (0.035) & 1.573 (0.030)  & --- &  \multicolumn{2}{c}{---}  \\ 
				& SCI (Debiased Lasso) & 99.4 & 2.013 (0.024) & 2.013 (0.024)  & --- &  \multicolumn{2}{c}{---}  \\ 
%				& Oracle (Bootstrap) & 89.8 & 0.651 (0.002) & 0.000 (0.000) & --- & \multicolumn{2}{c}{---}  \\ 	
				& Oracle & 95.5 & 0.649 (0.002) & 0.000 (0.000) & --- & \multicolumn{2}{c}{---}  \\ 	
%    Example 4-4 n=100, p=150, s0=15, N1000, r2000

				\bottomrule 
                \multicolumn{8}{l}{}\\ 
				\multicolumn{8}{l}{Note: SSCI is our Sparsified Simultaneous Confidence Intervals. We construct the SSCI using seven variable selection methods:}\\ 
				\multicolumn{8}{l}{SPSP based on solution paths of MCP (SPSP+MCP), SCAD (SPSP+SCAD), adaptive Lasso (SPSP+AdaLasso) and Lasso }\\ 
				\multicolumn{8}{l}{(SPSP+Lasso), SCAD with 10-fold cross-validation (SCAD+CV), MCP with 10-fold cross-validation (MCP+CV), adaptive }\\ 
				\multicolumn{8}{l}{Lasso with 10-fold cross-validation (AdaLasso+CV). For the alternative methods, we compare the simultaneous confidence}\\ 
				\multicolumn{8}{l}{intervals based on de-sparsified Lasso method (SCI (hdi)), the SCI based on double-selection approach of Lasso method }\\ 
				\multicolumn{8}{l}{(SCI (hdm)), and the SCI based on debiased Lasso method (SCI (Debiased Lasso)). We also compare SSCI with two reference}\\ 
				\multicolumn{8}{l}{methods. ``Oracle'' is simultaneous confidence intervals via the Bonferroni adjustment under the true model. $^\ast$ The SCI(hdm)}\\ 
				\multicolumn{8}{l}{method failed in the Example 8 when the p/n ratio is high.}\\ 
    
			\end{tabular}
		}
%	}
\end{table}

\bibliographystyle{plainnat}

\bibliography{bibliography}